\documentclass[prc,showpacs,showkeys,nofootinbib,superscriptaddress]{revtex4}
\usepackage{graphicx}
\usepackage{dcolumn}
\usepackage{epsfig}
\usepackage{bm}
\newcommand{\nj}[9]{\mbox{$\left \{ \begin{array}{ccc}#1 &#2 &#3 \\
#4 &#5 &#6 \\ #7 &#8 &#9 \end{array} \right \}$ }}

\newcommand{\tj}[6]{\mbox{$\left ( \begin{array}{ccc}#1 &#2 &#3\\
#4 &#5 &#6\end{array}\right )$ }}

\newcommand{\eq}{\begin{eqnarray}}
\newcommand{\en}{\end{eqnarray}}

\newcommand{\ba}[1]{\begin{eqnarray} \label{(#1)}}
\newcommand{\ea}{\end{eqnarray}}

\newcommand{\newc}{\newcommand}
\def\vbf{\mbox{\boldmath $\upsilon$}}
\def\bfs{\mbox{\boldmath $\sigma$}}
\newc{\lra}{\leftrightarrow}
\newc{\beq}{\begin{equation}}
\newc{\eeq}{\end{equation}}
\newc{\barr}{\begin{eqnarray}}
\newc{\earr}{\end{eqnarray}}
  \def\vbf{\mbox{\boldmath $\upsilon$}}
  \def\Sbf{\mbox{\boldmath $\Sigma$}}
\def\sbf{\mbox{\boldmath $\sigma$}}
\def\ebf{\mbox{\boldmath $\epsilon$}}
\def\gbf{\mbox{\boldmath $\gamma$}}

\begin{document}

\topmargin -0.50in
\title {Renormalization of the Spin-dependent  WIMP scattering off nuclei}
\author{P. C. Divari}
\affiliation{ Department of Physical Sciences and Applications,\\
Hellenic Army Academy,
 Vari 16673,Attica, Greece }

\author{J. D. Vergados \thanks{Vergados@cc.uoi.gr}}
\affiliation{Theoretical Physics Division, University of
Ioannina,\\ Ioannina, GR 451 10, Greece}

\begin{abstract}
 We study the amplitude  for the spin-dependent WIMP scattering off nuclei by including the leading
long-range two-body currents in the most important isovector contribution. We show that such effects are essentially independent
 of the target nucleus and, as a result, they can be treated as a mere renormalization of the effective nucleon cross section or,
 equivalently, of the corresponding effective coupling with values around 25$\%$.
\end{abstract}



\date{\today}

\maketitle

  \section{Introduction}
The combined MAXIMA-1 \cite{MAXIMA-1}, BOOMERANG \cite{BOOMERANG},
DASI \cite{DASI} and COBE/DMR Cosmic Microwave Background (CMB)
observations \cite{COBE} imply that the Universe is flat
 \cite{flat01}
and that most of the matter in
the Universe is Dark \cite{SPERGEL},  i.e. exotic. These results have been confirmed and improved
by the recent WMAP data \cite{WMAP06}. Combining
the data one finds:
$$\Omega_b=0.0456 \pm 0.0015, \quad \Omega _{\mbox{{\tiny CDM}}}=0.228 \pm 0.013 , \quad \Omega_{\Lambda}= 0.726 \pm 0.015~.$$
Since any ``invisible" non exotic component cannot possibly exceed $40\%$ of the above $ \Omega _{\mbox{{\tiny CDM}}}$
~\cite {Benne}, exotic (non baryonic) matter is required, i.e. there is room  cold dark matter candidates or WIMPs (Weakly Interacting Massive Particles).

Even though there exists firm indirect evidence for a halo of dark matter
in galaxies from the
observed rotational curves, see e.g. the review \cite{UK01}, it is essential to directly
detect
such matter.
The possibility of such detection, however, depends on the nature of the dark matter
constituents and their interactions.

Since the WIMP's are  expected to be
extremely non relativistic, with average kinetic energy $\langle T\rangle  \approx
50 \ {\rm keV} (m_{\mbox{{\tiny WIMP}}}/ 100 \ {\rm GeV} )$, they are not likely to excite the nucleus, even if they
are quite massive $m_{\mbox{{\tiny WIMP}}} > 100$ GeV.
Therefore they can be directly detected mainly via the recoiling of a nucleus
(A,Z) in elastic scattering. The event rate for such a process can
be computed from the following ingredients:
i) An effective Lagrangian at the elementary particle (quark)
level obtained in the framework of the prevailing particle theory.  The most popular scenario is in the context of supersymmetry.
 Here  the dark matter candidate is the
LSP (Lightest Supersymmetric Particle) \cite{ref2a,ref2b,ref2c,ref2,ELLROSZ,Gomez,ELLFOR}.
In this case the effective Lagrangian is constructed as described,
e.g., in
Refs.~\cite{ref2a,ref2b,ref2c,ref2,ELLROSZ,Gomez,ELLFOR,JDV96,JDV06a}.
At least till supersymmetry is discovered in the laboratory (LHC),
other approaches are also possible, such as, e.g., technibaryon
\cite{Nussinov92,GKS06}, mirror matter \cite{FLV72,Foot11},
Kaluza-Klein models with universal extra dimensions
\cite{ST02a,OikVerMou} ii) A well defined procedure for
transforming the amplitude obtained using the previous effective
Lagrangian, from the quark to the nucleon level. To achieve this
one needs a quark model for the nucleon, for the coherent mode
see, e.g., \cite{JDV06a,Dree00,Dree,Chen}. Similar  effects occur
in the case of the spin induced  matrix elements, iii) knowledge
of the relevant nuclear matrix elements \cite{Ress,DIVA00},
obtained with as reliable as possible many body nuclear wave
functions, iv) knowledge of the WIMP density in our vicinity and
its velocity distribution.

In the standard nuclear recoil experiments, first proposed more than 30 years ago \cite{GOODWIT},
one has to face the problem that the reaction of interest does not have a characteristic feature to distinguish it
from the background. So for the expected low counting rates the background is
a formidable problem. Some special features of the WIMP-nuclear interaction can be exploited to reduce the background problems, such as
 the modulation effect \cite{Druck},\cite{PSS88,GS93,RBERNABEI95,LS96,ABRIOLA98,HASENBALG98,JDV03,GREEN04,SFG06}
and  backward-forward asymmetry expected in directional experiments, i.e. experiments in which the direction of the
recoiling nucleus is also observed
 \cite{SPERGEL88}\cite{DRIFT,SHIMIZU03,KUDRY04,DRIFT2,GREEN05,Green06,KRAUSS,KRAUSS01,Alenazi08,Creswick010,Lisanti09,Giometal11}.

There exists a plethora of direct dark matter experiments  with the task of detecting  WIMP event rates
 for a variety of targets such as those employed in XENON10 \cite{XENON10}, XENON100 \cite{XENON100.11},
  XMASS \cite{XMASS09}, ZEPLIN \cite{ZEPLIN11}, PANDA-X \cite{PANDAX11}, LUX \cite{LUX11}, CDMS \cite{CDMS05},
  CoGENT \cite{CoGeNT11}, EDELWEISS \cite{EDELWEISS11}, DAMA \cite{DAMA1,DAMA11}, KIMS \cite{KIMS07} and PICASSO \cite{PICASSO09,PICASSO11}.

For the interpretation of the experimental data,  knowledge of the nuclear matrix elements is essential.
 In connection with nuclear structure aspects, in a series of calculations, e.g. in \cite{JDV03,JDV04,VF07}
 and references therein, it has been shown that for the coherent contribution, due to the scalar interaction,
 the inclusion of the nuclear form factor is important, especially in the case of relatively heavy targets.
  They also showed that the nuclear spin cross sections, which may be relevant for odd mass targets,
  are characterized by a single, i.e. essentially isospin independent, structure function and two static spin values, one for
   the  proton and one for the neutron, which depend on the target.


 In the present paper we will focus on the evaluation of the spin induced matrix elements.  In particular we will explore  the
currents for spin-dependent WIMP scattering off nuclei, in particular the most isovector channel. We will examine
i) the modification of the rates  at the one-body level by including a suitable nucleon form factor, which
 modifies somewhat the universality of the suitably normalized spin isospin structure functions and ii) include the leading
long-range two-body currents, which are predicted in chiral EFT \cite{MeGazSCH12}. We will show that the two body
 effects lead to a quenching of the isospin one spin matrix  elements. Since, however the resulting quenching factor is
 independent of the target, it can be absorbed into the corresponding nucleon cross section. We have not examined such
  effects on the isoscalar spin matrix  element since this mode, if present at all, is suppressed due to the quark structure of the nucleon.
  \section{The differential event rate}
  The expressions involving the differential are  fairly well known (see e.g. \cite{VergF12} for a brief review in our notation).
  For the reader's convenience we will list the main features here. The event rate can be cast in the form:
\beq
\left .\frac{d R}{ d E_R}\right |_A=\left .\frac{dR_0}{dE_R}\right |_A+\left .\frac{dR_1}{dE_R}\right |_A \cos{\alpha}
\eeq
where the first term represents the time averaged (non modulated) differential event rate, while the second  gives the
time dependent (modulated) one due to the motion of the Earth (see below). Furthermore for the coherent mode we get:
\barr
\left .\frac{d R_0}{ dE_R }\right |_A&=&\frac{\rho_{\chi}}{m_{\chi}}\,\frac{m_t}{A m_p}\, \left (\frac{\mu_r}{\mu_p} \right )^2\, \sqrt{<\upsilon^2>} \,\frac{1}{Q_0(A)} \sigma_A^{\mbox{\tiny{coh}}}\left .\left (\frac{d t}{du}\right ) \right |_{\mbox{ \tiny coh}}, \nonumber\\
\left .\frac{d R_1}{ d E_R}\right |_A&=&\frac{\rho_{\chi}}{m_{\chi}}\,\frac{m_t}{A m_p} \,\left (\frac{\mu_r}{\mu_p} \right )^2\, \sqrt{<\upsilon^2>} \,\frac{1}{Q_0(A)} \sigma_A^{\mbox{\tiny{coh}}}\left .\left (\frac{d h}{du}\right )  \right |_{\mbox{\tiny coh}},
\label{drdu}
\earr
 while for the spin dependent process we get:
\barr
\left .\frac{d R_0}{ dE_R }\right |_A&=&\frac{\rho_{\chi}}{m_{\chi}}\,\frac{m_t}{A m_p}\, \left (\frac{\mu_r}{\mu_p} \right )^2\, \sqrt{<\upsilon^2>} \,\frac{1}{Q_0(A)} \sigma_A^{\mbox{\tiny{spin}}}\left .\left (\frac{d t}{du}\right ) \right |_{\mbox{ \tiny spin}}, \nonumber\\
\left .\frac{d R_1}{ d E_R}\right |_A&=&\frac{\rho_{\chi}}{m_{\chi}}\,\frac{m_t}{A m_p} \,\left (\frac{\mu_r}{\mu_p} \right )^2\, \sqrt{<\upsilon^2>} \,\frac{1}{Q_0(A)} \sigma_A^{\mbox{\tiny{spin}}}\left .\left (\frac{d h}{du}\right )  \right |_{\mbox{\tiny spin}}
\label{drdus}
\earr
In the above expressions  $\mu_r$ ($\mu_p$) is the WIMP-nucleus (nucleon) reduced mass,  $A$ is the nuclear mass
number, $ m_{\chi}$ is the WIMP mass, $\rho(\chi)$ is the WIMP density in our vicinity, assumed to be 0.3 GeV cm$^{-3}$,
 and $m_t$ the mass of the target.
Furthermore one can show that
\beq
\sigma_A^{\mbox{\tiny{coh}}}=A^2 \sigma_N^{\mbox{\tiny{coh}}}, \quad \left .\left (\frac{d t}{d u}\right )\,\right |_{\mbox{\tiny coh}}=\sqrt{\frac{2}{3}}\, a^2 \,F^2(u)  \, \Psi_0(a \sqrt{u}),\quad \left .\left (\frac{d h}{d u}\right )\right |_{\mbox{\tiny coh}}=\sqrt{\frac{2}{3}}\, a^2 \,F^2(u) \,\Psi_1(a \sqrt{u})
\label{Eq.sigmacoh}
\eeq
with $ \sigma_N^{\mbox{\tiny{coh}}}$ the elementary nucleon cross section for  the coherent mode.
In the case of the spin  the expressions for static nuclear cross section $\sigma_A^{\mbox{\tiny{spin}}} $ will be given  below and:
  \beq
\left . \left (\frac{d t}{d u}\right )\right |_{\mbox{\tiny spin}}=\sqrt{\frac{2}{3}} \, a^2  \, F_{11}(u)   \,  \Psi_0(a \sqrt{u}),\quad \left .\left (\frac{d h}{d u}\right )\right |_{\mbox{\tiny spin}}=\sqrt{\frac{2}{3}}  \, a^2  \, F_{11}(u)  \, \Psi_1(a \sqrt{u}),
\label{Eq:Psi}
\eeq
The factor $\sqrt{2/3}$ is nothing but $\upsilon_0/\sqrt{\langle \upsilon ^2\rangle}$ since in Eq. (\ref{drdu}) $\sqrt{\langle \upsilon ^2\rangle}$
appears. In the above expressions  $a=(\sqrt{2} \mu_r b \upsilon_0)^{-1}$, $\upsilon_0$ the velocity of the sun around the center of the galaxy and
$b$ the nuclear harmonic oscillator size parameter characterizing the nuclear wave function.  $ u$ is the energy transfer $E_R$ in dimensionless
 units given by
\begin{equation}
 u=\frac{E_R}{Q_0(A)}~~,~~Q_{0}(A)=[m_pAb^2]^{-1}=40A^{-4/3}\mbox{ MeV}
\label{defineu}
\end{equation}
In the above expressions $F(u)$ is the nuclear form factor and $F_{11}(u)$ the isospin 1 spin response function.
Note that the parameter $a$ depends both on the WIMP mass, the target and the velocity distribution. Note also
that for a given energy transfer $E_R$ the quantity $u$ depends on $A$. \\
  The functions $\Psi_0(x)$ and $\Psi_1(x)$ arise from the WIMP velocity distribution, via function $g(\upsilon_{min},\upsilon_E(\alpha))$,
  which depends on  the minimum WIMP velocity for a given energy transfer, i.e.
\beq
\upsilon_{min}=\sqrt{\frac{A \,m_p \,E_R}{2 \, \mu^2_r}}
\eeq
For the M-B distribution in the local frame it is defined as follows:
\beq
g(\upsilon_{min},\upsilon_E(\alpha))=\frac{1}{\left (\sqrt{\pi}\upsilon_0 \right )^3}\int_{\upsilon_{min}}^{\upsilon_{max}}e^{-(\upsilon^2+2 \vbf . \vbf_E(\alpha)+\upsilon_E^2(\alpha))/\upsilon^2_0} \, \upsilon  \,  d\upsilon  \, d \Omega,\quad\upsilon_{max}=\upsilon_{esc}
\eeq
$\vbf_E(\alpha)$ is the velocity of the Earth, including the velocity of the Sun  around the galaxy, $\vbf_E(\alpha)= \epsilon_0({\hat\upsilon}_s+\delta \left (\sin{\alpha}{\hat x}-\cos{\alpha}\cos{\gamma}{\hat y}+\cos{\alpha}\sin{\gamma} {\hat \upsilon}_s\right ))$ with $\gamma\approx \pi/6$, $ {\hat \upsilon}_s$ a unit vector in the Sun's direction of motion, $\hat{x}$  a unit vector radially out of the galaxy in our position and  $\hat{y}={\hat \upsilon}_s\times \hat{x}$. $\delta=0.235$ is the velocity of the earth around the sun  divided  $\upsilon_0$ and $\alpha$ is the phase of the Earth ($\alpha=0$, around June 3nd)\footnote{One could, of course, make the time dependence of the rates due to the motion of the Earth more explicit by writing $\alpha \approx(6/5)\pi\left (2 (t/T)-1 \right )$, where $t/T$ is the fraction of the year.}. The above upper cut off value in the M-B is usually put in by hand. Such a cut off comes in naturally, however, in the case of velocity distributions obtained from the halo WIMP mass density in the Eddington approach \cite{VEROW06}, which, in certain models, resemble a M-B distribution \cite{JDV09}.
\\To obtain Eqs  (\ref{Eq.sigmacoh}) and (\ref{Eq:Psi}) we  expand $g(\upsilon_{min},\upsilon_E(\alpha))$ in powers of $\delta$,
keeping terms up to linear in $\delta \approx 0.135$. We found it convenient to  express all velocities in units of the Sun's
velocity $\upsilon_0$ to  obtain:
\beq
\upsilon_0 \, g(\upsilon_{min},\upsilon_E(\alpha))=\Psi_0(x)+\Psi_1(x)\cos{\alpha}, \quad x=\frac{\upsilon_{min}}{\upsilon_{0}}
\eeq
$\Psi_0(x)$ represents the quantity relevant for the average rate and $\Psi_1(x)$, which is proportional to $\delta$, represents
the effects of modulation. \\
In the case of a M-B distribution these functions take the following form:
\beq
\Psi_0(x)=\frac{1}{2}
  \left [\mbox{erf}(1-x)+\mbox{erf}(x+1)+\mbox{erfc}(1-y_{\mbox{\tiny{esc}}})+\mbox{erfc}(y_{\mbox{\tiny{esc}}}+1)-2 \right ]
  \label{Eq:Psi0MB}
\eeq
\barr
\Psi_1(x)&=&\frac{1}{2} ~\delta
   \left[\frac{ -\mbox{erf}(1-x)-\mbox{erf}(x+1)-\mbox{erfc}(1-y_{\mbox{\tiny{esc}}})-
   \mbox{erfc}(y_{\mbox{\tiny{esc}}}+1)}{2} \right . \nonumber\\
  && \left . +\frac{ e^{-(x-1)^2}}{\sqrt{\pi }}
   +\frac{
   e^{-(x+1)^2}}{\sqrt{\pi }}-\frac{ e^{-(y_{\mbox{\tiny{esc}}}-1)^2}}{\sqrt{\pi
   }}-\frac{ e^{-(y_{\mbox{\tiny{esc}}}+1)^2}}{\sqrt{\pi }}+1\right]
   \label{Eq:Psi1MB}
\earr
where erf$(x)$ and erfc$(x)$ are the error function and its complement, respectively, and $y_{esc}=\upsilon_{esc}/\upsilon_0$, $ 550 \leq\upsilon_{esc}\leq 660$ km/s.


  \section{The spin dependent WIMP-Nucleus scattering}
\begin{figure}[htb]
\begin{center}
{
\includegraphics[width=0.4\textwidth]{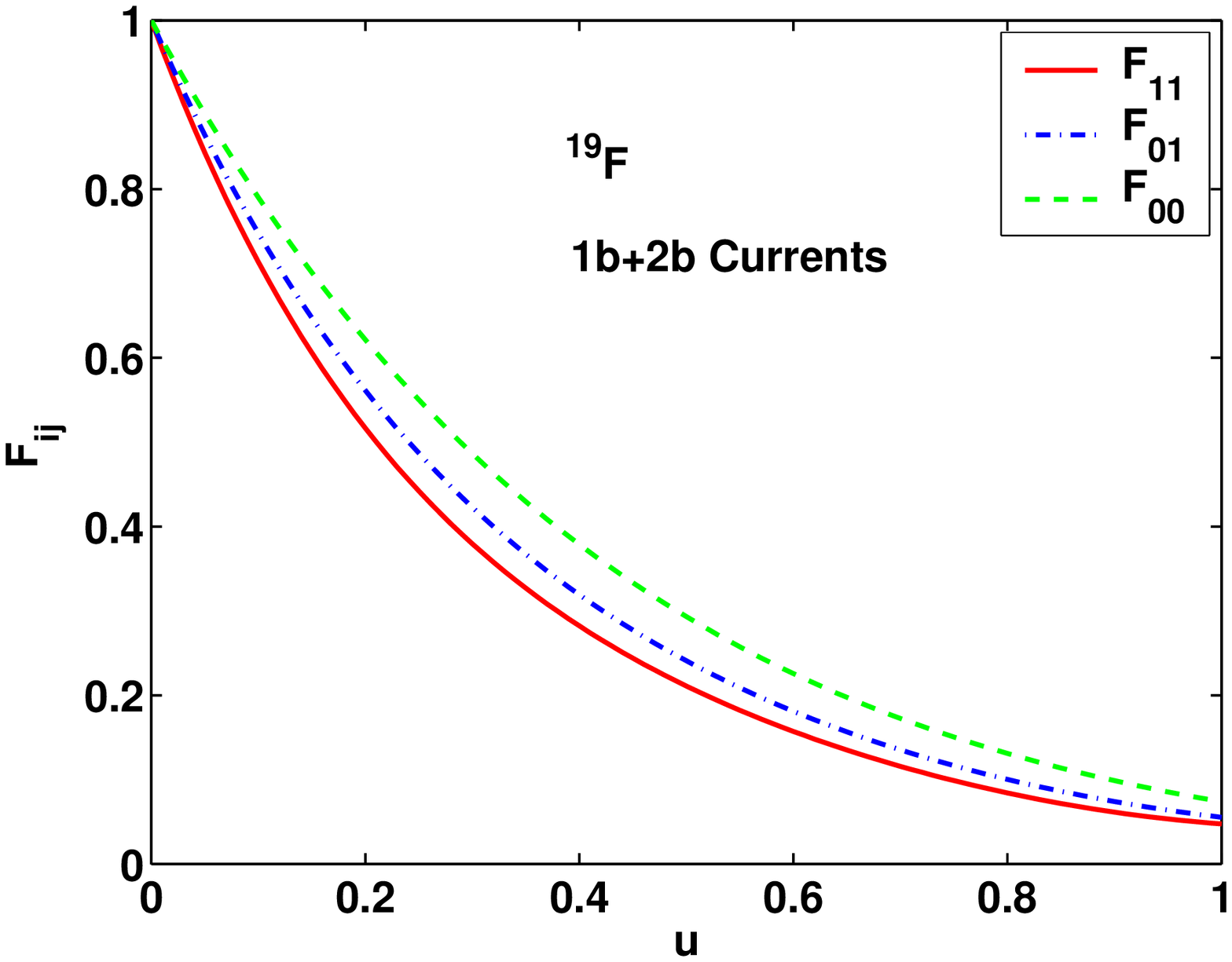}a)
} {
\includegraphics[width=0.4\textwidth]{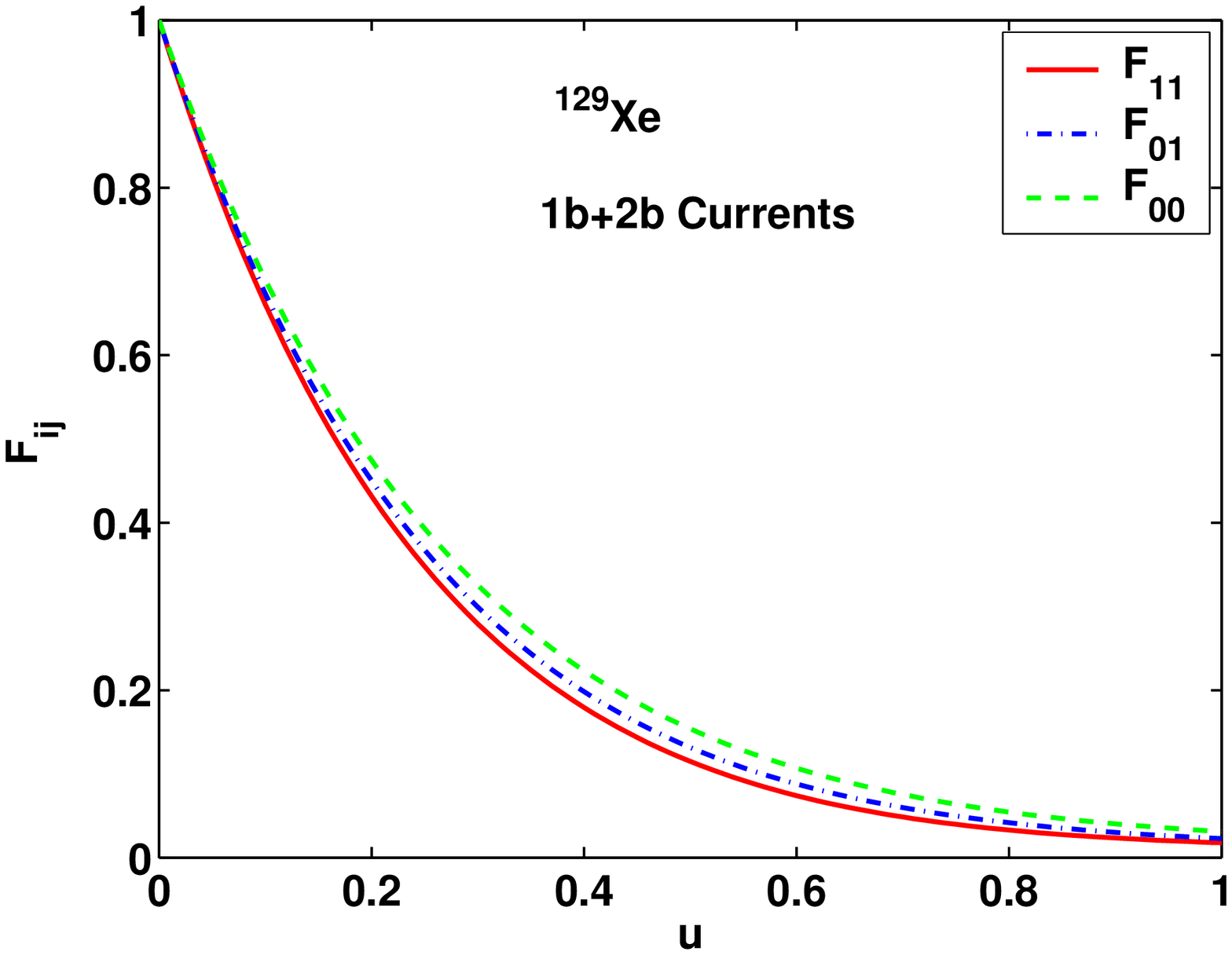}b)
}
{
\includegraphics[width=0.4\textwidth]{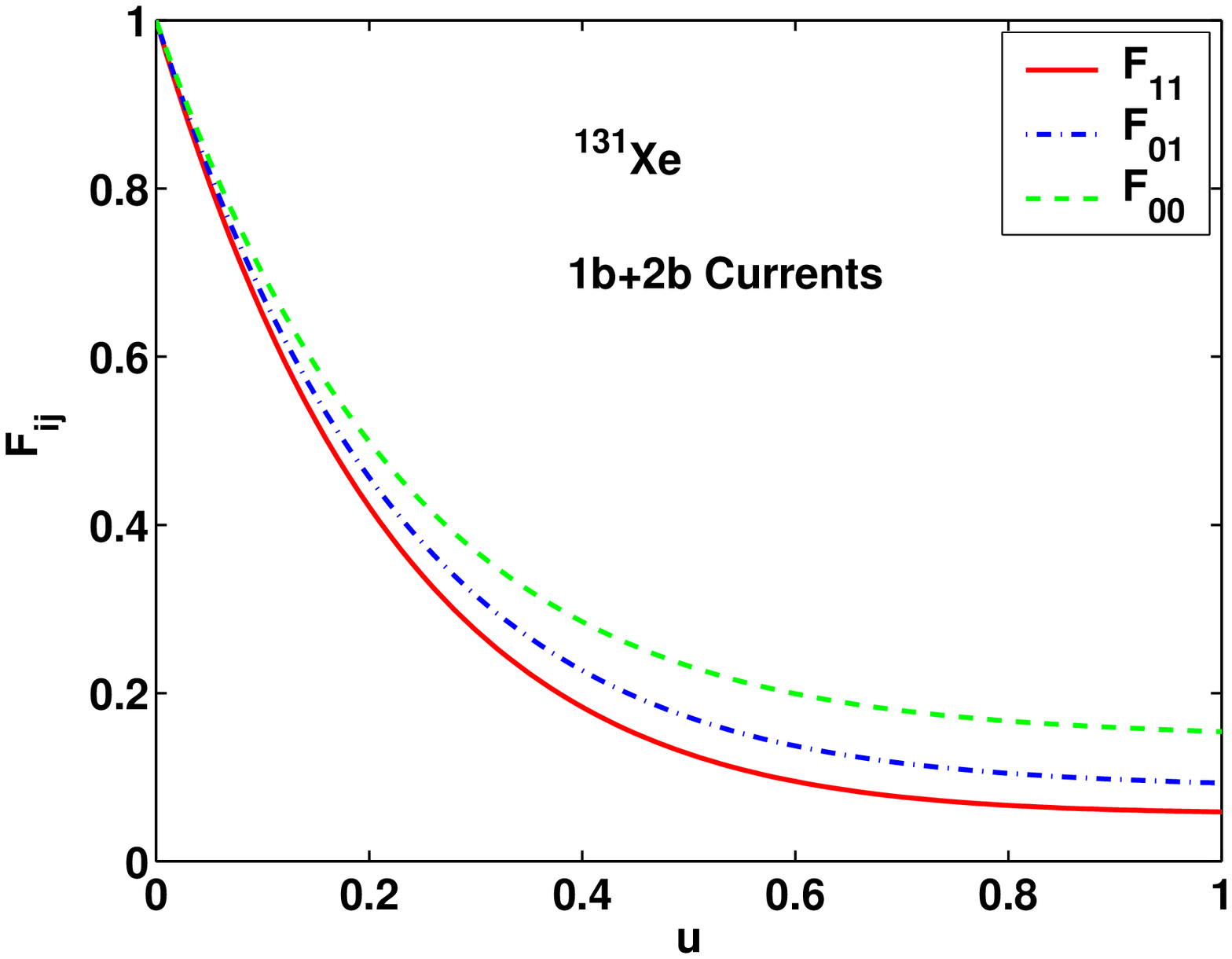}c)
}
\end{center}
\caption{(Color online) Structure factors $F_{00}$, $F_{11}$,
and $F_{01}$
 for $^{19}$F, $^{129}$Xe   and $^{131}$Xe isotopes  and as a function
of $u=q^2b^2/2$, for 1b+2b currents. No two body corrections were
considered for $F_{00}$. } \label{Fig:F129}
\end{figure}

  The spin dependent WIMP-nucleus elastic scattering cross section can be derived from an interaction which at the quark level can
  be cast in the form
  \begin{equation}
  {\cal L}=-\Lambda \frac{G_F}{\sqrt{2}} {\bf J} .\sum_q A_q \bar{\Psi}_q \gbf \gamma_5 \Psi_q
  \end{equation}
  Where the scale parameter $\Lambda$ as well as ${\bf J}$ depend on the particle model. Thus:
  \begin{enumerate}
  \item Supersymmetry,  see, e.g., a review in our notation \cite{CHIOS07} .\\
  \beq
  \Lambda=2(|C_{31}|^2
-|C_{41}|^2),\quad {\bf J}={\bar \chi}\gbf \gamma_5 \chi
  \eeq
  where $C_{31}$, $C_{41}$ are the Higgsino components of the WIMP, which in this case is the Lightest Supersymmeric Particle (LSP).
  The coefficients $A_q$ ca be read from the space component of the neutral current:
  \beq
J^Z_{\lambda} = - {\bar q} \gamma_{\lambda} \{ \frac {1}{3} sin^2
\theta_W - \Big[ \,\frac {1}{2} (1-\gamma_5) - sin^2 \theta_W
\Big]\tau_3 \} q,\quad A_0=0,\,A_1=\frac{1}{2},\,g_A^0(q)=0,\,
g_A^1(q)=-1.
  \label{eq:eg 13}
\eeq
\item Klauza Klein theories in models with Universal Extra Dimensions \cite{OikVerMou}.\\
In this case we encounter a number of cases depending on the nature of the WIMP, e,g.
\begin{itemize}
\item The WIMP is a gauge boson.\\
In this case:
\barr
\Lambda \approx \frac{8}{3} \tan^2{\theta_W} \left(\frac{m_W}{m_B}\right)^2\frac{m_p}{m_B}, \,
{\bf J}=(i \ebf^{'*}\times \ebf),\, A_u&=&\frac{17}{18},\,A_d=A_s=\frac{5}{18},\quad A_0=\frac{3}{2},\,A_1=\frac{2}{3}\nonumber\\
g_A^0(q)&=&g_A^1(q)=1
\earr
with $m_p$ the proton mass and $\ebf^{'}$ and $\ebf$ the polarization vectors of the initial and final WIMPs (the normalization
factors $1/\sqrt{m_B}$ entering each boson field in the expression of the cross section has already been absorbed into $\Lambda$).
\item The WIMP is heavy neutrino.\\
It is assumed the process proceeds via $Z$ exchange. A viable possibility is if it is a Majorana Fermion. In this case:
\beq
\Lambda=1, \quad {\bf J}=-\bar{\nu} \gbf \gamma_5\nu
\eeq
The coefficients $A_q$ are given again by the space component of the neutral current $ (A_0=0,\,  A_1=\frac{1}{2},\,g_A^0(q)=0,\,  g_A^1(q)=-1)$.
\end{itemize}
  \end{enumerate}
  In the present work we will not concern ourselves with any specific particle model, but  we will, instead, focus on the following points:
  \begin{enumerate}
  \item Going from the quark to the nucleon level, i.e.  constructing an effective transition operator at the nucleon level.  This involves:
 \begin{itemize}
 \item The modification of the currents due to the structure of the nucleon.\\
 The effective couplings $g^1_A$ ($g^0_A$) are obtained by multiplying the corresponding
elementary amplitudes obtained at the quark level, as given above, by suitable
renormalization factors $g^0_A$ and $g^1_A$ given in terms of the
quantities $\Delta q$ given by Ellis \cite{JELLIS},\cite{OikVerMou}, namely $\Delta u=0.78 \pm 0.02$, $\Delta d=-0.48\pm 0.02 $
and $\Delta s=-0.15 \pm 0.02.$, i.e.
\beq
g_A^0(N)=g_A^0(q)(\Delta u+\Delta d+\Delta s)=0.13g_A^0(q),\quad g_A^1(N)=g_A^1(q)(\Delta u-\Delta d)=1.26 g_A^1(q)
\eeq
The quantities $g_A^1(q)$ and $g_A^1(q)$ can be read from the hadronic current. For the neutral current  they are unity, so the
renormalization of the isovector component of the axial current is the usual one, while the isoscalar one is greatly suppressed,
 consistent with the EMC effect, i.e. the fact a tiny fraction of the spin of the quarks is coming from the spin of the quarks.
 It is for this reason that we started our discussion in the isospin basis. In the case of the boson WIMP in Kaluza-Klein
 theories one finds \cite{OikVerMou}:
\beq
g^0_A(N)=\frac{17}{18}\Delta u+\frac{5}{18}(\Delta d+\Delta s)=0.26,\quad g^1_A(N)=\frac{17}{18}\Delta u-\frac{5}{18}\Delta d=0.41
\eeq
 Thus in general,
barring very unusual circumstances at the quark level, the
isovector component is expected to be dominant.
 \item The nucleon form factor \cite{DIVA00,IJMPE}, which modifies the isovector axial current:
 \beq
 {\bf J}=\frac{g_A(q)}{g_A}\Sbf,\,\quad \Sbf=\bfs -\frac{\left ( \bfs.{\bf q}\right ){\bf q}}{q^2+m_{\pi}^2}
 \label{Eq:gAFf}
 \eeq
 If we choose as a $z$-axis the direction of momentum we find:
 \beq
 {\bf J}_m= \left \{ \begin{array}{l}\left (1-\frac{q^2}{q^2+m^2_{\pi}}\right )\bfs_m, \quad m=0\\ \bfs_{m}, \quad m=\pm1\end{array} \right.
 \eeq
Thus only the longitudinal component of the transition operator is modified by the inclusion of the nucleon form factor.
Thus, even at sufficiently high momentum transfers, only 1/3 of the differential rate will be affected.
 \item possible effects arising from possible exchange currents, which lead to effective 2-body contributions. We will
 exploit recent work on Chiral Effective Field Theory (EFT) \cite{MeGazSCH12}.
 \end{itemize}
  \item Construct the relative Spin Structure functions. \\
  For this purpose and for illustration purposes we will consider the simplest possible nuclear system, which in our
   opinion is the $A=19$ system \cite{DIVA00} currently employed as a target by the PICASSO collaboration. The study
   of other experimentally more interesting targets , see e.g. ref. \cite{PICASSO09}, can be found elsewhere \cite{MeGazSCH12}.
    \end{enumerate}
  The nuclear ME entering the WIMP-nucleus cross section takes the form:
   \beq
  |ME|^2 =a_1^2 S_{11}(u)+a_1 a_0 S_{01}(u) +a^2_{0}S_{00}(u),\quad  a_0=A_0 g^0_A(N),\quad a_1=A_1 g^1_A(N),
  \eeq
  i.e. the parameters  $a_0$ and $a_1$   depend on the  parameters $A_q$ as well as on  the quark model for the
  nucleon. The structure functions
  \beq
  F_{ij}(u)=\frac{S_{ij}(u)}{S_{ij}(0)},
  \eeq
 with the functions $ S_{ij}$ defined in the appendix,\footnote{The functions $S_{ij}$ for the Xe
  isotopes \cite{Ressa,Ressb} are normalized differently. The functions $F_{ij}$ extracted from them are
  independent of this normalization.}   are essentially independent of the isospin. Thus one can simplify
  the expression involving  the nuclear ME entering the WIMP-nucleus cross section and write it as follows:
 \beq
 |ME|^2 =\left (a_1^2  \Omega^2_1+a_1 a_0   \Omega_0 \Omega_1+a^2_{0} \Omega^2_0 \right) F_{11}(u),
  \label{Eq:MEsq}
  \eeq
 where  $\Omega_0$  and $\Omega_1$ are the  static nuclear  matrix elements involving  $\sbf$ for isospin 0 and 1 respectively defined by
 \beq
 \Omega_p=2 \sqrt{\frac{j+1}{j}}\prec[jj|S_p|jj\succ,\quad \Omega_n=2 \sqrt{\frac{j+1}{j}}\prec[jj|S_n|jj\succ,\quad \Omega_0=\Omega_p+\Omega_n\quad\Omega_1=\Omega_p-\Omega_n
 \eeq
  \\As it is common practice the nuclear physics parameters $\Omega_p$ and $\Omega_n$ (or $\Omega_0$ and $\Omega_1$) as well   $F_{11}$
   appear in the expression of the rate, while the parameters $\Lambda$, $a_1$ and $a_0$ can be absorbed into the nucleon cross sections
   $\sigma_p$ and $\sigma_n$ (or $\sigma_0(N)$ and $\sigma_1(N)$). Leaving aside the factor $F_{11}$, taken care of explicitly in Eq. (\ref{Eq:Psi}),
   the static nuclear cross section becomes
   \beq
  \sigma_A^{\mbox{\tiny{spin}}}=\frac{\left (  \Omega^2_1\sigma_1(N)+2\mbox{sign}(a_1 a_0)  \Omega_0 \Omega_1\sqrt{\sigma_0(N)\sigma_1(N)}+ \Omega^2_0 \sigma_0(N)\right)}{3}
  \label{Eq:nosf}
  \eeq
with
\beq
\sigma_0(N)=\Lambda^2 a_0^2\frac{G^2_F m_N^2}{2 \pi},\quad \sigma_1(N)=\Lambda^2 a_1^2\frac{G^2_F m_N^2}{2 \pi}
\eeq
and $m_N$ the nucleon mass.\\
The structure functions $F_{ij}$ obtained in this work are shown
in Fig. \ref{Fig:F129}(a), while those derived from a previous
work \cite{MeGazSCH12}  are exhibited in Fig. \ref{Fig:F129}(b)
and (c). The isospin behavior of the functions $F_{ij}$ in the
$A=129$ is what we expect. We do not understand the stronger
isospin dependence in the case of the $A=131$.

 Before proceeding
further we will interpret the results of the recent work
\cite{MeGazSCH12}, summarized below (see  section \ref{sec:EFT}),
 in our notation.  If we  define:
  \beq
 r_{ij}(0)=\frac{S_{ij}(0)}{\Omega_i \Omega_j}
 \eeq
 then
\begin{enumerate}
\item for the $^{129}$Xe isotope we take \\
$$\Omega_p=0.034,\quad \Omega_n=1.140,\quad \Omega_0=1.174,\quad \Omega_1=-1.105,$$
and
$$r_{00}(0)=r_{11}(0)=r_{01}(0)=\frac{1}{8 \pi}\simeq 0.038 \mbox{ (1-body) },$$
$$r_{00}(0)=\frac{1}{8 \pi},\quad r_{11}(0)=0.0240,$$
$$r_{01}(0)=0.0614  \mbox{ (one and two-body) }$$

\item for the $^{131}$Xe isotope we have \\
$$ \Omega_p=-0.023,\quad \Omega_n=-0.702,\quad \Omega_0=-0.776,\quad \Omega_1=0.679,$$
and
$$r_{00}(0)=r_{11}(0)=r_{01}(0)=\frac{1}{4 \pi}\simeq 0.079 \mbox{ (1-body) },$$
$$r_{00}(0)=\frac{1}{4 \pi},\quad r_{11}(0)=0.0487,\quad r_{01}(0)= 0.1155 \mbox{ (one and two-body) }$$
The above results have been obtained  from Eq. (16) as well as the
data of Table 1 of Ref.  \cite{MeGazSCH12}.

\item In the case of $^{19}$F we have \cite{DIVA00}:
 \beq
\Omega_p=1.644,\quad \Omega_n=-0.031,\quad
\Omega_0=1.614,\quad\Omega_1=1.675 \eeq The precise value of the
parameters $r_{ij}$ depends on the parameters of the model. For a
reasonable choice of these parameters we get: \barr
r_{00}(0)&=&r_{11}(0)=r_{01}(0)=1 \mbox{ (1-body) },\nonumber\\
r_{00}(0)&=&1,\quad r_{11}(0)=0.600,\quad
r_{01}(0)=0.775 \mbox{ (one and two-body) } \label{Eq:r11a} \earr
\end{enumerate}
In the presence of effective two body terms we can write: \beq
\left . r_{11}(0) \right |_{(1+2)b}=(1-\delta)^2 \left . r_{11}(0)
\right |_{1b},\quad \left . r_{01}(0) \right |_{(1+2)b}=(1-\delta)
\left . r_{01}(0) \right |_{1b} \eeq From the above analysis we
find: \beq \delta\approx0.229 \mbox{ (A=129)}, \quad\delta \approx
0.234 \mbox{ (A=131)}, \quad\delta \approx 0.235 \mbox{ (A=19)}
\eeq The corrections due to the presence of two body terms are
significant.
  From the above expressions, since $\delta$ is independent of A, one has two possibilities:
  \begin{enumerate}
  \item Write Eq. (\ref{Eq:nosf}) in the form
 \beq
   \sigma_A^{\mbox{\tiny{spin}}}=\frac{\left (\sigma_1 (1-\delta)^2 \Omega^2_1+\mbox{sign}(a_0 a_1) \sqrt{\sigma_1} \sqrt{\sigma_0} (1-\delta) \Omega_0 \Omega_1+\sigma_0 \Omega^2_0 \right)}{3} ,
  \label{Eq:MEsq2}
  \eeq
  where the nucleon parameters remain the same but the isovector nuclear spin  matrix elements are quenched by a factor of $(1-\delta)$.
  \item View the correction factor like an effective charge.\\
  In this case, due to the nuclear medium, the elementary nucleon amplitudes are modified:
   \beq
 a_0\rightarrow a_0,\quad a_1\rightarrow a_1(1-\delta),\quad \sigma_0\rightarrow \sigma_0,\quad \sigma_1\rightarrow \sigma_1(1-\delta)^2
 \eeq
but the nuclear spin matrix elements remain unchanged. In this case we recover the standard formula
  \beq
  \sigma_A^{\mbox{\tiny{spin}}} =\frac{\left (\sigma_1  \Omega^2_1+\mbox{sign}(a_0 a_1) \sqrt{\sigma_1} \sqrt{\sigma_0}  \Omega_0 \Omega_1+\sigma_0 \Omega^2_0 \right) ,
  \label{Eq:ME0sq2}}{3}
  \eeq
 \end{enumerate}
 The parameter $\delta$ depends on the assumed model as shown in table \ref{tab:delta}.\\
 One may express the above results in the proton neutron representation.  The above behavior of the form factors $F_{ij}$
  imply that the proton and neutron form factors are almost the same. If one retains the amplitudes unchanged one finds:
  \barr
 \sigma_A^{\mbox{\tiny{spin}}}&=&\frac{1}{3}\left[ \sigma_p \left (\Omega_p(1-\frac{\delta}{2})+\Omega_n \frac{\delta}{2} \right)^2+\sigma_n \left (\Omega_n(1-\frac{\delta}{2})+\Omega_p \frac{\delta}{2} \right)^2\right .\nonumber\\
  &+&\left . 2\mbox{sign}(a_p a_n)\left (\Omega_p \Omega_n\frac{1+(1-\delta)^2}{2}+(\Omega_p^2+\Omega_n^2)\frac{1-(1-\delta)^2}{4} \right )\right ]
  \earr
with
\beq
\sigma_p=\Lambda^2 a_p^2\frac{G^2_F m_N^2}{2 \pi},\quad \sigma_n=\Lambda^2 a_n^2\frac{G^2_F m_N^2}{2 \pi},\quad a_p=a_0+a_1,\, a_n=a_0-a_1
\eeq
In other words the two-body terms induce a small neutron (proton) component in those cases in which such a component
is negligible in the standard treatment with only one body terms.\\ On the other hand one can absorb the quenching into the
coupling constants, in which case the proton and the neutron cross sections get modified :
\beq
 a_p=a_0+a_1(1-\delta),\quad a_n=a_0-a_1(1-\delta),\quad \sigma_p\rightarrow\sigma_p(a_0+a_1(1-\delta))^2,\quad\sigma_n\rightarrow\sigma_n(a_0-a_1(1-\delta))^2,
\eeq
while the standard expression for the nuclear cross section is retained, i.e.
  \beq
 \sigma_A^{\mbox{\tiny{spin}}}=\frac{1}{3}\left[ \sigma_p \Omega_p^2+\sigma_n \Omega_n^2+ 2\mbox{sign}(a_p a_n) \sqrt{\sigma_p} \sqrt{\sigma_n} \Omega_p \Omega_n \right ]
  \eeq
\begin{table}
 \caption{The parameter $\delta$ entering the suppression
  factor $f=(1-\delta)$ for the axial isovector contribution
   for various values of the model parameters $\rho$, $I(\rho,P=0)$, $c_3$ and $c_4$.}
 \label{tab:delta}
 \begin{center}
 \begin{tabular}{||c|c|c|c|c||}
 \hline
 \hline
 $\rho$ & $I$ &   $c_3$ &  $c_4 $&  $ \delta $ \\
 \hline
  0.10&  0.58& -3.2&  5.40&        0.256\\
  \hline
  0.10 & 0.58 &-2.2&  4.40 &       0.203\\
  \hline
  0.10&  0.58& -3.40 & 3.40 &      0.189\\
  \hline
  0.10&  0.58& -2.40 & 2.40 &       0.136\\
  \hline
  0.10 & 0.58 &-4.78&  3.96&      0.233\\
  \hline
  0.10 & 0.58& -3.78&  2.96&       0.180\\
  \hline
  0.11 &0.59& -3.20&  5.40 &      0.287\\
  \hline
  0.11& 0.59 &-2.20 & 4.40  &     0.227\\
  \hline
  0.11 &0.59& -3.40  &3.40&     0.212\\
  \hline
  0.11& 0.59 &-2.40 & 2.40 &      0.152\\
  \hline
  0.11& 0.59& -4.78 & 3.96 &    0.261\\
  \hline
  0.11& 0.59 &-3.78&  2.96&      0.202\\
  \hline
  0.12 & 0.60& -3.20&  5.40&        0.318\\
  \hline
  0.12 & 0.60 &-2.20  &4.40  &      0.252\\
  \hline
  0.12&  0.60& -3.40 & 3.40 &      0.235\\
  \hline
  0.12 & 0.60& -2.4 0& 2.40 &       0.169\\
  \hline
  0.12 & 0.60 &-4.78 & 3.96&      0.290\\
  \hline
  0.12 & 0.60 &-3.78 & 2.96  &    0.224\\
  \hline
  \hline
  \end{tabular}
 \end{center}
 \end{table}

\section{The spin structure functions in the context of EFT}

  \label{sec:EFT}
  In this section we will employ the normalization of the structure functions even though, as shown above,
  they will be used only in extracting the relevant static spin matrix elements and in particular the quenching factors.
 Following the recent work of Ref. \cite{MeGazSCH12} we have included   the currents
for spin-dependent WIMP scattering off nuclei at the one-body
level as well as  the leading long-range two-body currents, which
are predicted in chiral EFT.
 The nuclear matrix elements (ME) entering the WIMP-nucleus cross section
  can be written as \cite{IJMPE}
  \begin{equation}
  \label{ME}
|ME|^2=\Big|<J||{\cal T}^{el \hspace{2pt}5}_{L}||J>\Big|^2+\Big
|<J||{\cal
 L}^{5}_{L}||J>\Big|^2,\hspace{4pt}L=odd
\end{equation}
where ${\cal T}^{el5}_{L}$ is the transverse electric and ${\cal
L}^{5}_{L}$ the longitudinal projections of the axial-currents.
Since the ground state of $^{19}$F is the $J=1/2^+$ then only the
$L=1$   multipole contributes.
  As consequence   the above operators
can be written in the isospin basis in  terms of the operators
$T^{(l,L)} =\sqrt{4\pi} j_{l}(q r_i) [Y^{l}(\hat{\bf r_i})
\,\times {\bm \sigma}]^L$ as
\begin{eqnarray}
 \label{eq1}
 {\cal L}^{5}_{L=1}(q)  =\frac{1}{\sqrt{
3}}\frac{1}{2}\frac{1}{\sqrt{4\pi}}\sum_{i=1}^A
\Big[a_0(\sqrt{2}\hspace{2pt}T^{(2,1)} +T^{(0,1)}) + a_1\tau_{i}^3
f_1(q)(\sqrt{2}\hspace{2pt}T^{(2,1)} +T^{(0,1)}) \Big ]
\end{eqnarray}
and
\begin{eqnarray}
  \label{eq2}
 {\cal T}^{el \hspace{2pt}5}_{L=1} (q) =\frac{1}{\sqrt{
3}}\frac{1}{2}\frac{1}{\sqrt{4\pi}}\sum_{i=1}^A
\Big[a_0(-\hspace{2pt}T^{(2,1)} +\sqrt{2}\hspace{2pt}T^{(0,1)}) +
 a_1\tau_{i}^3f_2(q)(-\hspace{2pt}T^{(2,1)}
+\sqrt{2}\hspace{2pt}T^{(0,1)}) \Big ]
\end{eqnarray}
where
\begin{eqnarray}
 \label{eq12aa}
  f_1(q)= 1 - \delta  - \frac{2 g_{\pi p n} F_\pi \, q^2}{2 m_N g_A
(m_\pi^2+q^2)}  -2 c_3 \, \frac{\rho}{F^2_\pi} \frac{q^2}{4
m^2_{\pi}+q^2}
\end{eqnarray}
and
\begin{eqnarray}  \label{eq11aa}
f_2(q)=  1 - \delta -2 \, \frac{q^2}{\Lambda_{A}^2}
\end{eqnarray}
The above operators are given in terms of the variable
$u=q^2b^2/2$, where  $q$ is the momentum transfer and $b$ the
harmonic-oscillator length.  Using the results of our previous
shell model calculation for $^{19}$F \cite{DIVA00} the parameter
$b$ has been taken the value 1.63 fm. The multi-particle matrix
elements of the above operators in the considered sd model space
are given in the Appendix.

The resulting 2b contribution to the axial-vector WIMP current has
been  included as a density-dependent renormalization
\cite{MeGazSCH12} $a_1 (1 + \delta)$, with
\begin{equation}
\label{delta1} \delta  \equiv  \frac{\rho}{F^2_\pi} \, I(\rho,P=0)
\biggl( \frac{1}{3} \, (2c_4-c_3) + \frac{1}{6 m_N} \biggr) \,.
\end{equation}
A  typical range of the densities in nuclei $\rho= 0.10...0.12
fm^{-3}$ has been taken into account. This leads to $I(\rho,P) =
0.64...0.66$, using the Fermi-gas mean-value $P^2 = 6 k_f^2/5$
where $k_f$ the Fermi momentum, and $I(\rho,P) = 0.58...0.60$ for
$P=0$. The  coupling constants $c_3$ and $c_4$ describing the
contributions of long-range one-pion-exchange as well as the
short-range parts in the two-body currents are  taken from Ref.
\cite{MeGazSCH11}.
\begin{figure}[htb]
\begin{center}
{
\includegraphics[width=1.0\textwidth]{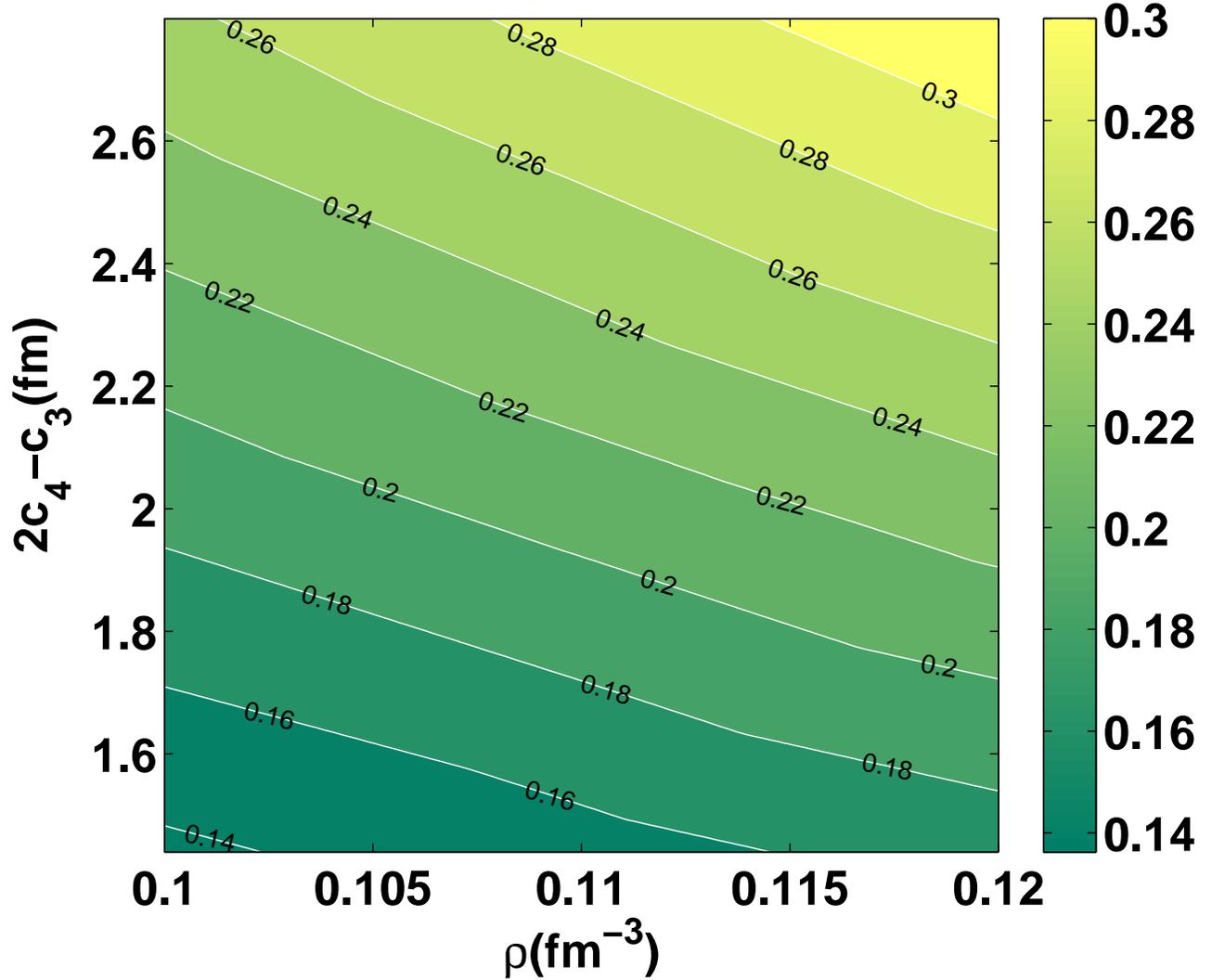}
}
{
}
\end{center}
\caption{(Color online) Contour plot of the  parameter $\delta$ as
a function of density $\rho$ and the difference $2c_4-c_3$ of the
parameters $c_3$ and  $c_4$.   The  curves in the plot correspond
to constant values of $\delta$ and lighter shading denotes the
increase of $\delta$. \label{cotour_delta}}
\end{figure}
In Fig. \ref{cotour_delta} the contour plot is shown for the
parameter $\delta$ as function of $\rho$ and $2c_4-c_3$. The
curves in the plot correspond to constant values of $\delta$.

\begin{figure}[htb]
\begin{center}
{
\includegraphics[width=0.4\textwidth]{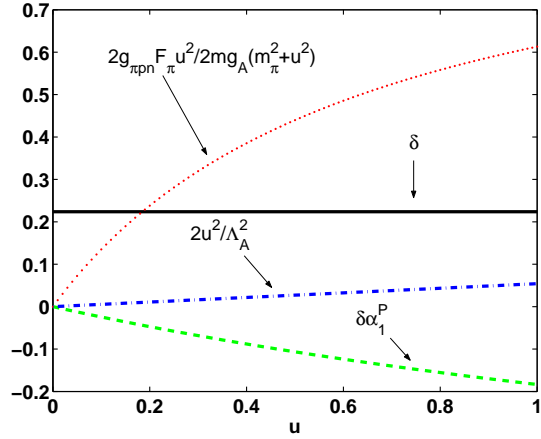}
}
\end{center}
\caption{(Color online) Form factors terms inserting in Eqs.
(\protect{\ref{eq12aa}}) and (\ref{eq11aa}). } \label{inspection}
\end{figure}

A second contribution $\delta \alpha^P_1$ in the pseudo-scalar
component of 2b current  has been included via a density and
momentum dependent modification:
\begin{equation}
\delta \alpha^P_1(q^2) \equiv 2 c_3 \, \frac{\rho}{F^2_\pi}
\frac{q^2}{4 m^2_{\pi}+q^2}
\end{equation}
where $m_{\pi}=138.04$~MeV and $F_{\pi}=92.4$~MeV. All the other
parameters in Eqs. (\ref{eq1}) and (\ref{eq2}) of the appendix
have been taken the values $\Lambda_{A}=1040$~MeV, $g_{\pi p
n}=13.05$, $g_A=1.27$. The first $q^2$-dependent term of $f_2(q)$
and $f_1(q)$ arise arise at the one-body (1b) currents, while the
second $q^2$-dependent term in $f_2(q)$ and the momentum
independent term $\delta$ are taken from the two body (2b)
currents. All the terms of the form factors $f_1(q)$ and $f_2(q)$
are presented in Fig. \ref{inspection}. As it is seen the first
$q^2$-dependent term of $f_2(q)$ is negligible. In Fig.
\ref{inspection} the corresponding average value  of the  two body
terms $\delta$ and $\delta \alpha^P_1$ is  presented.
\begin{table}[htb]
\centering \caption{ Fits to the structure factors $S_{00}$,
$S_{11}$ and $S_{01}$ for spin-dependent WIMP elastic scattering
off $^{19}$F nucleus, including 1b and 2b currents. The fitting
functions are given in terms of an exponential $e^{-u}$ multiplied
by a forth-order polynomial. The rows give the coefficients of the
$u^n$ terms in the polynomial.\label{fits}}
\begin{tabular*}{0.565\textwidth}{c|c|c|c|c|c}
\hline
\multicolumn{6}{c}{$^{19}$F} \\
\multicolumn{6}{c}{$u=q^2b^2/2 \,, \: b=1.63 \, {\rm fm}$} \\
\hline $e^{-u}\times$ & $S_{00}$ & $S_{11}$ (1b) & $S_{11}$
(1b+2b) &
$S_{01}$ (1b) & $S_{01}$ (1b+2b) \\
\hline
$1$ & $0.10386$ & $0.11122$ & $0.067215$ & $0.10756$ & $0.083473$ \\
$u$ & $-0.13848$ & $-0.26231$ & $-0.16378$ & $-0.20194$ & $-0.16079$ \\
$u^2$ & $0.06994$ & $0.36150$ & $0.23770$ & $0.20083$ & $0.16999$ \\
$u^3$ & $-0.01585$ & $-0.30632$ & $-0.20965$ & $-0.13217$ & $-0.11902$ \\
$u^4$ & $0.00136$ & $0.11021$ & $0.07718$ & $0.04167$ & $0.03888$ \\
\hline
\end{tabular*}
\end{table}
At finite momentum $p$ the structure functions $S_{00}$, $S_{11}$
and $S_{01}$ can be given in terms of a forth-order polynomial
multiplied by an exponential $e^{-u}$ (see Appendix). The results
of the fitting procedure are    given in Table \ref{fits}, while
the structure functions $S_{ij}$ for 1b and 1b+2b currents are
presented in Fig. \ref{inspection}.
\begin{figure}[htb]
\begin{center}
{
\includegraphics[width=0.4\textwidth]{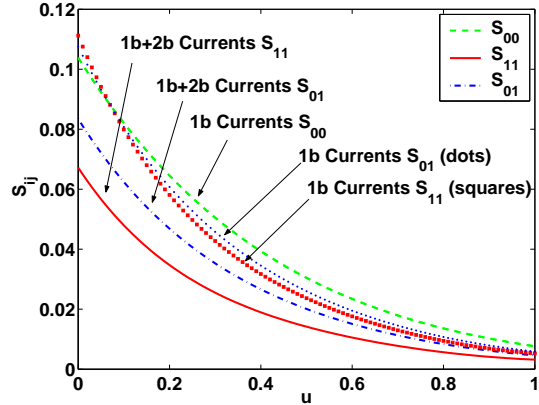}
}
\end{center}
\caption{(Color online)  Structure functions $S_{00}$,$S_{11}$ and
$S_{01}$
for $^{19}$F as a function of $u=q^2b^2/2$,  for one body(1b) and
one+two body(1b+2b) currents. No two body corrections  were
considered for $S_{00}$.
  \label{Fig:F19}}
\end{figure}

\section{Discussion}
In the present paper we studied  the evaluation of the spin
dependent nuclear matrix elements relevant for dark matter
searches with a variety of experimentally interesting targets (see
recent work  \cite{Cannoni11,Cannoni12} on such an analysis). We
focused on the isovector part of the axial current, which is the
most important for the spin induced cross section .

In particular, the effect on the suitably normalized  spin
structure functions $F_{ij}$ of the nucleon form factor has been
examined.
 We find that their isospin behavior of the spin structure functions  $F_{ij}$ is no longer universal.
 The isospin one modes fall a bit faster as a function of the energy transfer. This effect, however, is not
 very significant in the energy transfer of interest to dark matter searches (see Fig. \ref{Fig:F129}).
 In a recent analysis \cite{CVG11}, in which it is shown that it is possible to extract all three nucleon cross sections from the
 data (the coherent as well as the proton and neutron spin cross sections), the equality of the spin structure functions  $F_{ij}$ is
 important. In the case of the light nucleus $^{19}$F, in which the spin dependent rate maybe more important, we find that even the effect
 on the integrated structure functions $I_{ij}$, obtained by integrating  from zero to the maximum allowed energy transfer,  is small:
$$R_{01}=\frac{I_{01}}{I_{00}}= 0.965948 ,\quad R_{11}=\frac{I_{11}}{I_{00}}=0.936502$$
The effect will be smaller, if the additional $u$ dependence
arising from the velocity distribution, not discussed in this
work, is included. This  isospin independent function is also a
decreasing function of $u$, but its precise form depends on the
WIMP mass.


In addition to the standard one body contribution, we examined
the leading long-range two-body currents, which are predicted in
chiral EFT \cite{MeGazSCH12}.
 We confirm   the reduction of the isovector matrix elements, quenching factor, by as large as $15-25\%$.
 The precise value is dependent on model parameters, not precisely known. We have found, however, that this affects only the static
 spin ME and it is independent of the target nucleus. Thus this effect can be absorbed as a multiplicative factor either in the
 isovector nuclear matrix elements or better still in the effective coupling parameters, which depend on  the assumed particle model,
 which anyway are going to be derived from the experimental data, if and when they become available.
In summary  the standard analysis of the data previously employed is not seriously affected by such effects.

\section{Appendix}

Since the ground state of $^{19}$F is the $J=1/2^+$ then only the
$L=1$   multipole contributes (Eq. (\ref{ME})). In our treatment
we will separate the couplings from the nuclear operator.
  As consequence    the multi-particle reduced matrix elements
  of the
  ${\cal L}^{5}_{L=1}$ and ${\cal T}^{el \hspace{2pt}5}_{L=1}$ operators can
  be written as
\beq <J||{\cal L}^{5}_{L=1}||J> =a_0<J||{\cal L}^{5,0}_{L=1}||J>
+a_1<J||{\cal L}^{5,1}_{L=1}||J>
 \label{eqboth}
 \eeq
 where
 \beq
 <J||{\cal L}^{5,0}_{L=1}||J> =\frac{1}{\sqrt{
3}}\frac{1}{2}\frac{1}{\sqrt{4\pi}}\sum_{\lambda\rho}
a_{\lambda\rho}(J) (\sqrt{2}\hspace{2pt}<\lambda||T^{(2,1)}||\rho>
+<\lambda||T^{(0,1)}||\rho>)
 \label{eqbotha}
 \eeq
 \beq
 <J||{\cal L}^{5,1}_{L=1}||J> =\frac{1}{\sqrt{
3}}\frac{1}{2}\frac{1}{\sqrt{4\pi}} f_1(q)\sum_{\lambda\rho}
b_{\lambda\rho}(J)(\sqrt{2}\hspace{2pt}<\lambda||T^{(2,1)}||\rho>
+<\lambda||T^{(0,1)}||\rho>)
 \label{eqbothb}
 \eeq
 with
\begin{eqnarray}  \label{(eq12)}
  f_1(q)= 1 - \delta  - \frac{2 g_{\pi p n} F_\pi \, q^2}{2 m_N g_A
(m_\pi^2+q^2)}  -2 c_3 \, \frac{\rho}{F^2_\pi} \frac{q^2}{4
m^2_{\pi}+q^2}
\end{eqnarray}
 We also write
 \beq
 <J||{\cal T}^{el \hspace{2pt}5}_{L=1}||J>=a_0 <J||{\cal T}^{el \hspace{2pt}(5,0)}_{L=1}||J>+a_1 <J||{\cal T}^{el \hspace{2pt}(5,1)}_{L=1}||J>
 \eeq
 with
 \beq
 <J||{\cal T}^{el \hspace{2pt}5,0}_{L=1}||J>=\frac{1}{\sqrt{
3}}\frac{1}{2}\frac{1}{\sqrt{4\pi}}\sum_{\lambda\rho}
a_{\lambda\rho}(J) (-<\lambda||T^{(2,1)}||\rho>
+\sqrt{2}\hspace{2pt}<\lambda||T^{(0,1)}||\rho>)
 \label{eqbothc}
 \eeq
 and
 \beq
 <J||{\cal T}^{el \hspace{2pt}5,1}_{L=1}||J>= \frac{1}{\sqrt{
3}}\frac{1}{2}\frac{1}{\sqrt{4\pi}}f_2(q)\sum_{\lambda\rho}
b_{\lambda\rho}(J)(-<\lambda||T^{(2,1)}||\rho>
+\sqrt{2}\hspace{2pt}<\lambda||T^{(0,1)}||\rho>)
 \label{eqbothd}
 \eeq
with
\begin{eqnarray}  \label{(eq11)}
f_2(q)=  1 - \delta -2 \, \frac{q^2}{\Lambda_{A}^2}
\end{eqnarray}

The indices $\lambda$ and $\rho$  run over the single particle
orbits of the chosen model space. According to Ref. \cite{DIVA00}
this model space is the sd one. The quantities $a_{\lambda\rho}$
and $b_{\lambda\rho}$ are essentially products of the one body
coefficients of fractional parentage (CFP) for the isoscalar  and
isovector part of the operator respectively. They depend, of
course, on the specific interaction and the model space used (see
Ref. \cite{DIVA00}).

As it is known the reduced matrix elements of the $T^{(l,j)}
=\sqrt{4\pi} j_{l}(q r) [Y^{l}(\hat{\bf r}) \,\times {\bm
\sigma}]^j$ operator is written as
\begin{equation}
\label{Teq} T^{(l,j)}= {\cal
A}_{\lambda\rho}^{(l,j)}\int_0^{\infty}
 R_{n_\lambda
l_\lambda}(r)j_{l}(q r)R_{n_\rho l_\rho}(r)r^2dr
\end{equation}
where
\begin{equation}
\label{Aeq} {\cal A}_{\lambda\rho}^{(l,j)}=(-1)^{l_\lambda}\hat
{l_\lambda}\hat {l_\rho}\hat {j_\lambda}\hat {j_\rho}\hat {l}\hat
{j}
 \nj{l_\lambda}{
1/2}{j_\lambda}{l_\rho}{1/2}{j_\lambda}{l}{1}{j}\tj{l_\lambda}{l}{l_\rho}{0}{0}{0}\sqrt{6}
\end{equation}
Therefore the matrix elements (\ref{eqbotha}),
(\ref{eqbothb}),(\ref{eqbothc}) and (\ref{eqbothd}) are written as
\begin{eqnarray}
 \label{eq101}
<J||{\cal L}^{5,0}_{L=1}||J> =\frac{1}{\sqrt{
3}}\frac{1}{2}\frac{1}{\sqrt{4\pi}}
\Big(\sqrt{2}\hspace{2pt}T_0^{(2,1)}(u) +T_0^{(0,1)}(u)\Big)
\end{eqnarray}
\begin{eqnarray}
 \label{eq101a}
<J||{\cal L}^{5,1}_{L=1}||J> =\frac{1}{\sqrt{
3}}\frac{1}{2}\frac{1}{\sqrt{4\pi}}
f_1(u)\Big(\sqrt{2}\hspace{2pt}T_1^{(2,1)}(u) +T_1^{(0,1)}(u)\Big)
\end{eqnarray}
\begin{eqnarray}
  \label{eq102}
<J||{\cal T}^{el \hspace{2pt}5,0}_{L=1}||J> =\frac{1}{\sqrt{
3}}\frac{1}{2}\frac{1}{\sqrt{4\pi}}
\Big(-\hspace{2pt}T_0^{(2,1)}(u)
+\sqrt{2}\hspace{2pt}T_0^{(0,1)}(u)\Big)
\end{eqnarray}
\begin{eqnarray}
  \label{eq102a}
<J||{\cal T}^{el \hspace{2pt}5,1}_{L=1}||J> =\frac{1}{\sqrt{
3}}\frac{1}{2}\frac{1}{\sqrt{4\pi}}
 f_2(u) \Big(-\hspace{2pt}T_1^{(2,1)}(u)
+\sqrt{2}\hspace{2pt}T_1^{(0,1)}(u)\Big)
\end{eqnarray}
with
\begin{eqnarray}  \label{eq3}
T_0^{(0,1)}(u)=\Big [ \frac{1}{6}\Big( \frac{2}{5}A_1+A_2\Big)u^2
- \frac{2}{3}\Big(A_1+A_2\Big)u+A_1+A_2\Big ]e^{-u/2}
\end{eqnarray}
\begin{eqnarray}  \label{eq4}
T_0^{(2,1)}(u)=\Big [
\Big(-\frac{1}{15}A_1'+\frac{1}{6}\sqrt{\frac{2}{5}}A_2'\Big)u^2+\Big
(\frac{7}{15}A_1'-\frac{2}{3}\sqrt{\frac{2}{5}}A_2'\Big)u\Big]e^{-u/2}
\end{eqnarray}
\begin{eqnarray}  \label{eq5}
T_1^{(0,1)}(u)=\Big [ \frac{1}{6}\Big( \frac{2}{5}B_1+B_2\Big)u^2
- \frac{2}{3}\Big(B_1+B_2\Big)u+B_1+B_2\Big ]e^{-u/2}
\end{eqnarray}
\begin{eqnarray}  \label{eq6}
T_1^{(2,1)}(u)=\Big [
\Big(-\frac{1}{15}B_1'+\frac{1}{6}\sqrt{\frac{2}{5}}B_2'\Big)u^2+\Big
(\frac{7}{15}B_1'-\frac{2}{3}\sqrt{\frac{2}{5}}B_2'\Big)u\Big]e^{-u/2}
\end{eqnarray}
The $T_0^{(l,j)}(u)$  part of the operator represents the
isoscalar part while $T_1^{(l,j)}(u)$ the isovector one.
 Using the
results  of our previous  shell model calculation for $^{19}$F
\cite{DIVA00} the coefficients
$A_1$,$A_2$,$B_1$,$B_2$,$A_1'$,$A_2'$, $B_1'$ and $B_2'$ have been
taken the values
\begin{eqnarray}  \label{eq7}
 A_1=1.19407, \ A_2=1.09082, \ B_1=1.3695, \ B_2=0.999
\end{eqnarray}
and
\begin{eqnarray}  \label{eq8}
 A_1'=-0.004, \ A_2'=-0.0587, \ B_1'=0.0774, \ B_2'=0.06269
\end{eqnarray}
The above operators are given in terms of the variable
$u=q^2b^2/2$, where  $q$ is the momentum transfer and $b$ the
harmonic-oscillator length.

According to the above matrix elements the structure functions are
written
\begin{eqnarray}  \label{s00}
 S_{00}&=&8 \pi <J||{\cal L}^{5,0}_{L=1}||J>^2+<J||{\cal T}^{el
 \hspace{2pt}5,0}_{L=1}||J>^2\nonumber \\
 &=&\frac{1}{2}\Bigg [ \Big(T_0^{(2,1)}(u)\Big)^2+\Big(T_0^{(0,1)}(u)\Big)^2  \Bigg ]
\end{eqnarray}
\begin{eqnarray}  \label{s11}
 S_{11}&=8 \pi &<J||{\cal L}^{5,1}_{L=1}||J>^2+<J||{\cal T}^{el
 \hspace{2pt}5,1}_{L=1}||J>^2\nonumber \\
 &=&\frac{1}{6}\Bigg [\Big(2f_1^2(u)+f_2^2(u)\Big) \Big(T_1^{(2,1)}(u)\Big)^2+\Big(f_1^2(u)+2f_2^2(u)\Big) \Big(T_1^{(0,1)}(u)\Big)^2\nonumber \\
 &+&2\sqrt{2} T_1^{(2,1)}(u)T_1^{(0,1)}(u)\Big(f_1^2(u)-f_2^2(u)\Big)\Bigg ]
\end{eqnarray}
and
\begin{eqnarray}  \label{s01}
 S_{01}&=&8 \pi <J||{\cal L}^{5,0}_{L=1}||J> <J||{\cal L}^{5,1}_{L=1}||J> +
 <J||{\cal T}^{el \hspace{2pt}5,0}_{L=1}||J><J||{\cal T}^{el \hspace{2pt}5,1}_{L=1}||J>\nonumber \\
 &=&\frac{1}{6}\Bigg [\Big(2f_1(u)+f_2(u)\Big)T_0^{(2,1)}(u)T_1^{(2,1)}(u)+\sqrt{2}\Big(f_1(u)-f_2(u)\Big)T_0^{(2,1)}(u)T_1^{(0,1)}(u)\nonumber \\
 &+&\sqrt{2}\Big(f_1(u)-f_2(u)\Big)T_0^{(0,1)}(u)T_1^{(2,1)}(u)+\Big(f_1(u)+2f_2(u)\Big)T_0^{(0,1)}(u)T_1^{(0,1)}(u)\Bigg]
\end{eqnarray}


 \section*{References}

\begin{thebibliography}{82}
\expandafter\ifx\csname natexlab\endcsname\relax\def\natexlab#1{#1}\fi
\expandafter\ifx\csname bibnamefont\endcsname\relax
  \def\bibnamefont#1{#1}\fi
\expandafter\ifx\csname bibfnamefont\endcsname\relax
  \def\bibfnamefont#1{#1}\fi
\expandafter\ifx\csname citenamefont\endcsname\relax
  \def\citenamefont#1{#1}\fi
\expandafter\ifx\csname url\endcsname\relax
  \def\url#1{\texttt{#1}}\fi
\expandafter\ifx\csname urlprefix\endcsname\relax\def\urlprefix{URL }\fi
\providecommand{\bibinfo}[2]{#2}
\providecommand{\eprint}[2][]{\url{#2}}

\bibitem[{MAX()}]{MAXIMA-1}
\bibinfo{note}{S. Hanary {\it et al}: {\it Astrophys. J.} {\bf 545}, L5
  (2000);\\ J.H.P Wu {\it et al}: {\it Phys. Rev. Lett.} {\bf 87}, 251303
  (2001);\\ M.G. Santos {\it et al}: {\it Phys. Rev. Lett.} {\bf 88}, 241302
  (2002)}.

\bibitem[{BOO()}]{BOOMERANG}
\bibinfo{note}{P. D. Mauskopf {\it et al}: {\it Astrophys. J.} {\bf 536}, L59
  (2002);\\ S. Mosi {\it et al}: {\it Prog. Nuc.Part. Phys.} {\bf 48}, 243
  (2002);\\ S. B. Ruhl {\it al}, astro-ph/0212229 and references therein.}

\bibitem[{DAS()}]{DASI}
\bibinfo{note}{N. W. Halverson {\it et al}: {Astrophys. J.} {\bf 568}, 38
  (2002)\\ L. S. Sievers {\it et al}: astro-ph/0205287 and references therein.}

\bibitem[{\citenamefont{Smoot and {\it et
  al}~(COBE~Collaboration)}(1992)}]{COBE}
\bibinfo{author}{\bibfnamefont{G.~F.} \bibnamefont{Smoot}} \bibnamefont{and}
  \bibinfo{author}{\bibnamefont{{\it et al}~(COBE~Collaboration)}},
  \bibinfo{journal}{Astrophys. J.} \textbf{\bibinfo{volume}{396}},
  \bibinfo{pages}{L1} (\bibinfo{year}{1992}).

\bibitem[{\citenamefont{Jaffe and {\it et al}}(2001)}]{flat01}
\bibinfo{author}{\bibfnamefont{A.~H.} \bibnamefont{Jaffe}} \bibnamefont{and}
  \bibinfo{author}{\bibnamefont{{\it et al}}}, \bibinfo{journal}{Phys. Rev.
  Lett.} \textbf{\bibinfo{volume}{86}}, \bibinfo{pages}{3475}
  (\bibinfo{year}{2001}).

\bibitem[{\citenamefont{Spergel and {\it et al}}(2003)}]{SPERGEL}
\bibinfo{author}{\bibfnamefont{D.~N.} \bibnamefont{Spergel}} \bibnamefont{and}
  \bibinfo{author}{\bibnamefont{{\it et al}}}, \bibinfo{journal}{Astrophys. J.
  Suppl.} \textbf{\bibinfo{volume}{148}}, \bibinfo{pages}{175}
  (\bibinfo{year}{2003}).

\bibitem[{\citenamefont{Spergel et~al.}(2007)}]{WMAP06}
\bibinfo{author}{\bibfnamefont{D.}~\bibnamefont{Spergel}} \bibnamefont{et~al.},
  \bibinfo{journal}{Astrophys. J. Suppl.} \textbf{\bibinfo{volume}{170}},
  \bibinfo{pages}{377} (\bibinfo{year}{2007}),
  \bibinfo{note}{[arXiv:astro-ph/0603449v2]}.

\bibitem[{\citenamefont{Bennett and {\it et al}}(1995)}]{Benne}
\bibinfo{author}{\bibfnamefont{D.~P.} \bibnamefont{Bennett}} \bibnamefont{and}
  \bibinfo{author}{\bibnamefont{{\it et al}}}, \bibinfo{journal}{Phys. Rev.
  Lett.} \textbf{\bibinfo{volume}{74}}, \bibinfo{pages}{2867}
  (\bibinfo{year}{1995}).

\bibitem[{\citenamefont{Ullio and Kamioknowski}(2001)}]{UK01}
\bibinfo{author}{\bibfnamefont{P.}~\bibnamefont{Ullio}} \bibnamefont{and}
  \bibinfo{author}{\bibfnamefont{M.}~\bibnamefont{Kamioknowski}},
  \bibinfo{journal}{JHEP} \textbf{\bibinfo{volume}{0103}}, \bibinfo{pages}{049}
  (\bibinfo{year}{2001}).

\bibitem[{\citenamefont{Bottino and {\it et al.}}(1997)}]{ref2a}
\bibinfo{author}{\bibfnamefont{A.}~\bibnamefont{Bottino}} \bibnamefont{and}
  \bibinfo{author}{\bibnamefont{{\it et al.}}}, \bibinfo{journal}{Phys. Lett B}
  \textbf{\bibinfo{volume}{402}}, \bibinfo{pages}{113} (\bibinfo{year}{1997}).

\bibitem[{\citenamefont{Arnowitt and Nath}(1995)}]{ref2b}
\bibinfo{author}{\bibfnamefont{R.}~\bibnamefont{Arnowitt}} \bibnamefont{and}
  \bibinfo{author}{\bibfnamefont{P.}~\bibnamefont{Nath}},
  \bibinfo{journal}{Phys. Rev. Lett.} \textbf{\bibinfo{volume}{74}},
  \bibinfo{pages}{4592} (\bibinfo{year}{1995}).

\bibitem[{\citenamefont{Arnowitt and Nath}(1996)}]{ref2c}
\bibinfo{author}{\bibfnamefont{R.}~\bibnamefont{Arnowitt}} \bibnamefont{and}
  \bibinfo{author}{\bibfnamefont{P.}~\bibnamefont{Nath}},
  \bibinfo{journal}{Phys. Rev. D} \textbf{\bibinfo{volume}{54}},
  \bibinfo{pages}{2374} (\bibinfo{year}{1996}), \bibinfo{note}{hep-ph/9902237}.

\bibitem[{ref()}]{ref2}
\bibinfo{note}{A. Bottino {\it et al.}, {\it Phys. Lett B} {\bf 402}, 113
  (1997).\\ R. Arnowitt. and P. Nath, {\it Phys. Rev. Lett.} {\bf 74}, 4592
  (1995); {\it Phys. Rev. D} {\bf 54}, 2374 (1996); hep-ph/9902237;\\ V. A.
  Bednyakov, H.V. Klapdor-Kleingrothaus and S.G. Kovalenko, {\it Phys. Lett. B}
  {\bf 329}, 5 (1994).}

\bibitem[{\citenamefont{Ellis and Roszkowski}(1992)}]{ELLROSZ}
\bibinfo{author}{\bibfnamefont{J.}~\bibnamefont{Ellis}} \bibnamefont{and}
  \bibinfo{author}{\bibfnamefont{L.}~\bibnamefont{Roszkowski}},
  \bibinfo{journal}{Phys. Lett. B} \textbf{\bibinfo{volume}{283}},
  \bibinfo{pages}{252} (\bibinfo{year}{1992}).

\bibitem[{Gom()}]{Gomez}
\bibinfo{note}{M. E. G\'{o}mez and J. D. Vergados, {\it Phys. Lett. B} {\bf
  512} , 252 (2001); hep-ph/0012020.\\ M. E. G\'{o}mez, G. Lazarides and
  Pallis, C., Phys. Rev.D {\bf 61}, 123512 (2000) and {\it Phys. Lett. B} {\bf
  487}, 313 (2000).}

\bibitem[{ELL()}]{ELLFOR}
\bibinfo{note}{J. Ellis, and R. A. Flores, {\it Phys. Lett. B} {\bf 263}, 259
  (1991); {\it Phys. Lett. B} {\bf 300}, 175 (1993); {\it Nucl. Phys. B} {\bf
  400}, 25 (1993)}.

\bibitem[{\citenamefont{Vergados}(1996)}]{JDV96}
\bibinfo{author}{\bibfnamefont{J.~D.} \bibnamefont{Vergados}},
  \bibinfo{journal}{J. of Phys. G} \textbf{\bibinfo{volume}{22}},
  \bibinfo{pages}{253} (\bibinfo{year}{1996}).

\bibitem[{\citenamefont{Vergados}(2007)}]{JDV06a}
\bibinfo{author}{\bibfnamefont{J.~D.} \bibnamefont{Vergados}},
  \bibinfo{journal}{Lect. Notes Phys.} \textbf{\bibinfo{volume}{720}},
  \bibinfo{pages}{69} (\bibinfo{year}{2007}), \bibinfo{note}{hep-ph/0601064}.

\bibitem[{\citenamefont{Nussinov}(1992)}]{Nussinov92}
\bibinfo{author}{\bibfnamefont{S.}~\bibnamefont{Nussinov}},
  \bibinfo{journal}{Phys. Lett. B} \textbf{\bibinfo{volume}{279}},
  \bibinfo{pages}{111} (\bibinfo{year}{1992}).

\bibitem[{\citenamefont{Gudnason et~al.}(2006)\citenamefont{Gudnason, Kouvaris,
  and Sannino}}]{GKS06}
\bibinfo{author}{\bibfnamefont{S.~B.} \bibnamefont{Gudnason}},
  \bibinfo{author}{\bibfnamefont{C.}~\bibnamefont{Kouvaris}}, \bibnamefont{and}
  \bibinfo{author}{\bibfnamefont{F.}~\bibnamefont{Sannino}},
  \bibinfo{journal}{Phys. Rev. D} \textbf{\bibinfo{volume}{74}},
  \bibinfo{pages}{095008} (\bibinfo{year}{2006}),
  \bibinfo{note}{arXiv:hep-ph/0608055}.

\bibitem[{\citenamefont{Foot et~al.}(1991)\citenamefont{Foot, Lew, and
  Volkas}}]{FLV72}
\bibinfo{author}{\bibfnamefont{R.}~\bibnamefont{Foot}},
  \bibinfo{author}{\bibfnamefont{H.}~\bibnamefont{Lew}}, \bibnamefont{and}
  \bibinfo{author}{\bibfnamefont{R.~R.} \bibnamefont{Volkas}},
  \bibinfo{journal}{Phys. Lett. B} \textbf{\bibinfo{volume}{272}},
  \bibinfo{pages}{676} (\bibinfo{year}{1991}).

\bibitem[{\citenamefont{Foot}(2011)}]{Foot11}
\bibinfo{author}{\bibfnamefont{R.}~\bibnamefont{Foot}}, \bibinfo{journal}{Phys.
  Lett. B} \textbf{\bibinfo{volume}{703}}, \bibinfo{pages}{7}
  (\bibinfo{year}{2011}), \bibinfo{note}{[arXiv:1106.2688]}.

\bibitem[{\citenamefont{Servant and Tait}(2003)}]{ST02a}
\bibinfo{author}{\bibfnamefont{G.}~\bibnamefont{Servant}} \bibnamefont{and}
  \bibinfo{author}{\bibfnamefont{T.~M.~P.} \bibnamefont{Tait}},
  \bibinfo{journal}{Nuc. Phys. B} \textbf{\bibinfo{volume}{650}},
  \bibinfo{pages}{391} (\bibinfo{year}{2003}).

\bibitem[{\citenamefont{Oikonomou et~al.}(2007)\citenamefont{Oikonomou,
  Vergados, and Moustakidis}}]{OikVerMou}
\bibinfo{author}{\bibfnamefont{V.}~\bibnamefont{Oikonomou}},
  \bibinfo{author}{\bibfnamefont{J.}~\bibnamefont{Vergados}}, \bibnamefont{and}
  \bibinfo{author}{\bibfnamefont{C.~C.} \bibnamefont{Moustakidis}},
  \bibinfo{journal}{Nuc. Phys.} \textbf{\bibinfo{volume}{B 773}},
  \bibinfo{pages}{19} (\bibinfo{year}{2007}).

\bibitem[{Dre({\natexlab{a}})}]{Dree00}
\bibinfo{note}{A. Djouadi and M. K. Drees, {\it Phys. Lett. B} {\bf 484}, 183
  (2000); S. Dawson, {\it Nucl. Phys. B} {\bf 359}, 283 (1991); M. Spira {it et
  al}, {\it Nucl. Phys.} {\bf B453}, 17 (1995).}

\bibitem[{Dre({\natexlab{b}})}]{Dree}
\bibinfo{note}{M. Drees and M. M. Nojiri, {\it Phys. Rev. D} {\bf 48}, 3843
  (1993); {\it Phys. Rev. D} {\bf 47}, 4226 (1993).}

\bibitem[{Che()}]{Chen}
\bibinfo{note}{T. P. Cheng, {\it Phys. Rev. D} {\bf 38}, 2869 (1988); H-Y.
  Cheng, {\it Phys. Lett. B} {\bf 219}, 347 (1989).}

\bibitem[{Res()}]{Ress}
\bibinfo{note}{M. T. Ressell {\it et al.}, {\it Phys. Rev. D} {\bf 48}, 5519
  (1993); M.T. Ressell and D. J. Dean, Phys. Rev. C {\bf 56}, 535 (1997).}

\bibitem[{\citenamefont{Divari et~al.}(2000)\citenamefont{Divari, Kosmas,
  Vergados, and Skouras}}]{DIVA00}
\bibinfo{author}{\bibfnamefont{P.~C.} \bibnamefont{Divari}},
  \bibinfo{author}{\bibfnamefont{T.~S.} \bibnamefont{Kosmas}},
  \bibinfo{author}{\bibfnamefont{J.~D.} \bibnamefont{Vergados}},
  \bibnamefont{and} \bibinfo{author}{\bibfnamefont{L.~D.}
  \bibnamefont{Skouras}}, \bibinfo{journal}{Phys. Rev. C}
  \textbf{\bibinfo{volume}{61}}, \bibinfo{pages}{054612}
  (\bibinfo{year}{2000}).

\bibitem[{\citenamefont{Goodman and Witten}(1985)}]{GOODWIT}
\bibinfo{author}{\bibfnamefont{M.~W.} \bibnamefont{Goodman}} \bibnamefont{and}
  \bibinfo{author}{\bibfnamefont{E.}~\bibnamefont{Witten}},
  \bibinfo{journal}{Phys. Rev. D} \textbf{\bibinfo{volume}{31}},
  \bibinfo{pages}{3059} (\bibinfo{year}{1985}).

\bibitem[{\citenamefont{Drukier et~al.}(1986)\citenamefont{Drukier, Freeze, and
  Spergel}}]{Druck}
\bibinfo{author}{\bibfnamefont{A.}~\bibnamefont{Drukier}},
  \bibinfo{author}{\bibfnamefont{K.}~\bibnamefont{Freeze}}, \bibnamefont{and}
  \bibinfo{author}{\bibfnamefont{D.}~\bibnamefont{Spergel}},
  \bibinfo{journal}{Phys. Rev. D} \textbf{\bibinfo{volume}{33}},
  \bibinfo{pages}{3495} (\bibinfo{year}{1986}).

\bibitem[{\citenamefont{Primack et~al.}(1988)\citenamefont{Primack, Seckel, and
  Sadoulet}}]{PSS88}
\bibinfo{author}{\bibfnamefont{J.~R.} \bibnamefont{Primack}},
  \bibinfo{author}{\bibfnamefont{D.}~\bibnamefont{Seckel}}, \bibnamefont{and}
  \bibinfo{author}{\bibfnamefont{B.}~\bibnamefont{Sadoulet}},
  \bibinfo{journal}{Ann. Rev. Nucl. Part. Sci.} \textbf{\bibinfo{volume}{38}},
  \bibinfo{pages}{751} (\bibinfo{year}{1988}).

\bibitem[{\citenamefont{Gabutti and Schmiemann}(1993)}]{GS93}
\bibinfo{author}{\bibfnamefont{A.}~\bibnamefont{Gabutti}} \bibnamefont{and}
  \bibinfo{author}{\bibfnamefont{K.}~\bibnamefont{Schmiemann}},
  \bibinfo{journal}{Phys. Lett. B} \textbf{\bibinfo{volume}{308}},
  \bibinfo{pages}{411} (\bibinfo{year}{1993}).

\bibitem[{\citenamefont{Bernabei}(1995)}]{RBERNABEI95}
\bibinfo{author}{\bibfnamefont{R.}~\bibnamefont{Bernabei}},
  \bibinfo{journal}{Riv. Nouvo Cimento} \textbf{\bibinfo{volume}{18 (5)}},
  \bibinfo{pages}{1} (\bibinfo{year}{1995}).

\bibitem[{\citenamefont{Lewin and Smith}(1996)}]{LS96}
\bibinfo{author}{\bibfnamefont{J.~D.} \bibnamefont{Lewin}} \bibnamefont{and}
  \bibinfo{author}{\bibfnamefont{P.~F.} \bibnamefont{Smith}},
  \bibinfo{journal}{Astropart. Phys.} \textbf{\bibinfo{volume}{6}},
  \bibinfo{pages}{87} (\bibinfo{year}{1996}).

\bibitem[{\citenamefont{Abriola et~al.}(1999)}]{ABRIOLA98}
\bibinfo{author}{\bibfnamefont{D.}~\bibnamefont{Abriola}} \bibnamefont{et~al.},
  \bibinfo{journal}{Astropart. Phys.} \textbf{\bibinfo{volume}{10}},
  \bibinfo{pages}{133} (\bibinfo{year}{1999}),
  \bibinfo{note}{arXiv:astro-ph/9809018}.

\bibitem[{\citenamefont{Hasenbalg}(1998)}]{HASENBALG98}
\bibinfo{author}{\bibfnamefont{F.}~\bibnamefont{Hasenbalg}},
  \bibinfo{journal}{Astropart. Phys.} \textbf{\bibinfo{volume}{9}},
  \bibinfo{pages}{339} (\bibinfo{year}{1998}),
  \bibinfo{note}{arXiv:astro-ph/9806198}.

\bibitem[{\citenamefont{Vergados}(2003)}]{JDV03}
\bibinfo{author}{\bibfnamefont{J.~D.} \bibnamefont{Vergados}},
  \bibinfo{journal}{Phys. Rev. D} \textbf{\bibinfo{volume}{67}},
  \bibinfo{pages}{103003} (\bibinfo{year}{2003}),
  \bibinfo{note}{hep-ph/0303231}.

\bibitem[{\citenamefont{Green}(2003)}]{GREEN04}
\bibinfo{author}{\bibfnamefont{A.}~\bibnamefont{Green}},
  \bibinfo{journal}{Phys. Rev. D} \textbf{\bibinfo{volume}{68}},
  \bibinfo{pages}{023004} (\bibinfo{year}{2003}), \bibinfo{note}{ibid: D ${\bf
  69}$ (2004) 109902; arXiv:astro-ph/0304446}.

\bibitem[{\citenamefont{Savage et~al.}(2006)\citenamefont{Savage, Freese, and
  Gondolo}}]{SFG06}
\bibinfo{author}{\bibfnamefont{C.}~\bibnamefont{Savage}},
  \bibinfo{author}{\bibfnamefont{K.}~\bibnamefont{Freese}}, \bibnamefont{and}
  \bibinfo{author}{\bibfnamefont{P.}~\bibnamefont{Gondolo}},
  \bibinfo{journal}{Phys. Rev. D} \textbf{\bibinfo{volume}{74}},
  \bibinfo{pages}{043531} (\bibinfo{year}{2006}),
  \bibinfo{note}{arXiv:astro-ph/0607121}.

\bibitem[{\citenamefont{Spergel}(1988)}]{SPERGEL88}
\bibinfo{author}{\bibfnamefont{D.}~\bibnamefont{Spergel}},
  \bibinfo{journal}{Phys. Rev. D} \textbf{\bibinfo{volume}{37}},
  \bibinfo{pages}{1353} (\bibinfo{year}{1988}).

\bibitem[{DRI()}]{DRIFT}
\bibinfo{note}{The NAIAD experiment B. Ahmed {\it et al}, Astropart. Phys. {\bf
  19} (2003) 691; hep-ex/0301039\\ B. Morgan, A. M. Green and N. J. C. Spooner,
  Phys. Rev. D {\bf 71} (2005) 103507; astro-ph/0408047.}

\bibitem[{\citenamefont{Shimizu et~al.}(2003)\citenamefont{Shimizu, Minoa, and
  Inoue}}]{SHIMIZU03}
\bibinfo{author}{\bibfnamefont{Y.}~\bibnamefont{Shimizu}},
  \bibinfo{author}{\bibfnamefont{M.}~\bibnamefont{Minoa}}, \bibnamefont{and}
  \bibinfo{author}{\bibfnamefont{Y.}~\bibnamefont{Inoue}},
  \bibinfo{journal}{Nuc. Instr. Meth. A} \textbf{\bibinfo{volume}{496}},
  \bibinfo{pages}{347} (\bibinfo{year}{2003}).

\bibitem[{KUD()}]{KUDRY04}
\bibinfo{note}{V.A. Kudryavtsev, Dark matter experiments at Boulby mine,
  astro-ph/0406126.}

\bibitem[{\citenamefont{Morgan et~al.}(2005)\citenamefont{Morgan, Green, and
  Spooner}}]{DRIFT2}
\bibinfo{author}{\bibfnamefont{B.}~\bibnamefont{Morgan}},
  \bibinfo{author}{\bibfnamefont{A.~M.} \bibnamefont{Green}}, \bibnamefont{and}
  \bibinfo{author}{\bibfnamefont{N.~J.~C.} \bibnamefont{Spooner}},
  \bibinfo{journal}{Phys. Rev. D} \textbf{\bibinfo{volume}{71}},
  \bibinfo{pages}{103507} (\bibinfo{year}{2005}), \bibinfo{note}{;
  astro-ph/0408047}.

\bibitem[{\citenamefont{Morgan and Green}(2005)}]{GREEN05}
\bibinfo{author}{\bibfnamefont{B.}~\bibnamefont{Morgan}} \bibnamefont{and}
  \bibinfo{author}{\bibfnamefont{A.~M.} \bibnamefont{Green}},
  \bibinfo{journal}{Phys. Rev. D} \textbf{\bibinfo{volume}{72}},
  \bibinfo{pages}{123501} (\bibinfo{year}{2005}).

\bibitem[{\citenamefont{Green and Morgan}(2007)}]{Green06}
\bibinfo{author}{\bibfnamefont{A.~M.} \bibnamefont{Green}} \bibnamefont{and}
  \bibinfo{author}{\bibfnamefont{B.}~\bibnamefont{Morgan}},
  \bibinfo{journal}{Astropart. Phys.} \textbf{\bibinfo{volume}{27}},
  \bibinfo{pages}{142} (\bibinfo{year}{2007}), \bibinfo{note}{[ arXiv:0707.1488
  (astrp-ph)]}.

\bibitem[{\citenamefont{Copi et~al.}(1999)\citenamefont{Copi, Heo, and
  Krauss}}]{KRAUSS}
\bibinfo{author}{\bibfnamefont{C.}~\bibnamefont{Copi}},
  \bibinfo{author}{\bibfnamefont{J.}~\bibnamefont{Heo}}, \bibnamefont{and}
  \bibinfo{author}{\bibfnamefont{L.}~\bibnamefont{Krauss}},
  \bibinfo{journal}{Phys. Lett. B} \textbf{\bibinfo{volume}{461}},
  \bibinfo{pages}{43} (\bibinfo{year}{1999}).

\bibitem[{\citenamefont{Copi and Krauss}(2001)}]{KRAUSS01}
\bibinfo{author}{\bibfnamefont{C.}~\bibnamefont{Copi}} \bibnamefont{and}
  \bibinfo{author}{\bibfnamefont{L.}~\bibnamefont{Krauss}},
  \bibinfo{journal}{Phys. Rev. D} \textbf{\bibinfo{volume}{63}},
  \bibinfo{pages}{043507} (\bibinfo{year}{2001}).

\bibitem[{\citenamefont{Alenazi and Gondolo}(2008)}]{Alenazi08}
\bibinfo{author}{\bibfnamefont{A.}~\bibnamefont{Alenazi}} \bibnamefont{and}
  \bibinfo{author}{\bibfnamefont{P.}~\bibnamefont{Gondolo}},
  \bibinfo{journal}{Phys. Rev. D} \textbf{\bibinfo{volume}{77}},
  \bibinfo{pages}{043532} (\bibinfo{year}{2008}).

\bibitem[{Cre()}]{Creswick010}
\bibinfo{note}{R.J. Creswick and S. Nussinov and F.T. Avignone III, arXiv:
  1007.0214 [astro-ph.IM]}.

\bibitem[{Lis()}]{Lisanti09}
\bibinfo{note}{Lisanti and J.G. Wacker, arXiv: 0911.1997 [hep-ph]}.

\bibitem[{Gio()}]{Giometal11}
\bibinfo{note}{F. Mayet {\it et al}, Directional detection of dark matter,
  arXiv:1001.2983 (astro-ph.IM)}.

\bibitem[{XEN()}]{XENON10}
\bibinfo{note}{J. Angle {\it et al}, arXiv:1104.3088 [hep-ph]}.

\bibitem[{\citenamefont{Aprile et~al.}(2011)}]{XENON100.11}
\bibinfo{author}{\bibfnamefont{E.}~\bibnamefont{Aprile}} \bibnamefont{et~al.},
  \bibinfo{journal}{Phys. Rev. Lett.} \textbf{\bibinfo{volume}{107}},
  \bibinfo{pages}{131302} (\bibinfo{year}{2011}),
  \bibinfo{note}{arXiv:1104.2549v3 [astro-ph.CO]}.

\bibitem[{\citenamefont{Abe et~al.}(2009)}]{XMASS09}
\bibinfo{author}{\bibfnamefont{K.}~\bibnamefont{Abe}} \bibnamefont{et~al.},
  \bibinfo{journal}{Astropart. Phys.} \textbf{\bibinfo{volume}{31}},
  \bibinfo{pages}{290} (\bibinfo{year}{2009}), \bibinfo{note}{arXiv:v3
  [physics.ins-det]0809.4413v3 [physics.ins-det]}.

\bibitem[{\citenamefont{Ghag et~al.}(2011)}]{ZEPLIN11}
\bibinfo{author}{\bibfnamefont{C.}~\bibnamefont{Ghag}} \bibnamefont{et~al.},
  \bibinfo{journal}{Astropar. Phys.} \textbf{\bibinfo{volume}{35}},
  \bibinfo{pages}{76} (\bibinfo{year}{2011}), \bibinfo{note}{arXiv:1103.0393
  [astro-ph.CO]}.

\bibitem[{PAN()}]{PANDAX11}
\bibinfo{note}{See, e.g., Kaixuan Ni, Proceedings of the Dark Side of the
  Universe, DSU2011, Beijing, 9/27/2011}.

\bibitem[{LUX()}]{LUX11}
\bibinfo{note}{D.C. Malling {it et al}, arXiv:1110.0103((astro-ph.IM))}.

\bibitem[{\citenamefont{Akerib et~al.}(2006)}]{CDMS05}
\bibinfo{author}{\bibfnamefont{D.}~\bibnamefont{Akerib}} \bibnamefont{et~al.},
  \bibinfo{journal}{Phys. Rev. Lett.} \textbf{\bibinfo{volume}{96}},
  \bibinfo{pages}{011302} (\bibinfo{year}{2006}),
  \bibinfo{note}{arXiv:astro-ph/0509259 and arXiv:astro-ph/0509269}.

\bibitem[{\citenamefont{Aalseth et~al.}(2011)}]{CoGeNT11}
\bibinfo{author}{\bibfnamefont{C.}~\bibnamefont{Aalseth}} \bibnamefont{et~al.},
  \bibinfo{journal}{Phys. Rev. Lett.} \textbf{\bibinfo{volume}{106}},
  \bibinfo{pages}{131301} (\bibinfo{year}{2011}), \bibinfo{note}{coGeNT
  collaboration arXiv:10002.4703 [astro-ph.CO]}.

\bibitem[{\citenamefont{Armengaud et~al.}(2011)}]{EDELWEISS11}
\bibinfo{author}{\bibfnamefont{E.}~\bibnamefont{Armengaud}}
  \bibnamefont{et~al.}, \bibinfo{journal}{Phys. Lett. B}
  \textbf{\bibinfo{volume}{702}}, \bibinfo{pages}{329} (\bibinfo{year}{2011}),
  \bibinfo{note}{arXiv:1103.4070v3 [astro-ph.CO]}.

\bibitem[{\citenamefont{Bernabei and Others}(2008)}]{DAMA1}
\bibinfo{author}{\bibfnamefont{R.}~\bibnamefont{Bernabei}} \bibnamefont{and}
  \bibinfo{author}{\bibnamefont{Others}}, \bibinfo{journal}{Eur. Phys. J. C}
  \textbf{\bibinfo{volume}{56}}, \bibinfo{pages}{333} (\bibinfo{year}{2008}),
  \bibinfo{note}{[DAMA Collaboration]; [arXiv:0804.2741 [astro-ph]]}.

\bibitem[{DAM()}]{DAMA11}
\bibinfo{note}{P. Belli {\it et al}, arXiv:1106.4667 [astro-ph.GA]}.

\bibitem[{\citenamefont{Lee et~al.}(2007)}]{KIMS07}
\bibinfo{author}{\bibfnamefont{H.~S.} \bibnamefont{Lee}} \bibnamefont{et~al.},
  \bibinfo{journal}{Phys.Rev.Lett.} \textbf{\bibinfo{volume}{99}},
  \bibinfo{pages}{091301} (\bibinfo{year}{2007}),
  \bibinfo{note}{arXiv:0704.0423[astro-ph]}.

\bibitem[{\citenamefont{Archambault et~al.}(2009)}]{PICASSO09}
\bibinfo{author}{\bibfnamefont{S.}~\bibnamefont{Archambault}}
  \bibnamefont{et~al.}, \bibinfo{journal}{Phys. Lett. B}
  \textbf{\bibinfo{volume}{682}}, \bibinfo{pages}{185} (\bibinfo{year}{2009}),
  \bibinfo{note}{collaboration PICASSO, arXiv:0907.0307 [astro-ex]}.

\bibitem[{\citenamefont{Archambault et~al.}(2011)}]{PICASSO11}
\bibinfo{author}{\bibfnamefont{S.}~\bibnamefont{Archambault}}
  \bibnamefont{et~al.}, \bibinfo{journal}{New J. Phys.}
  \textbf{\bibinfo{volume}{13}}, \bibinfo{pages}{043006}
  (\bibinfo{year}{2011}), \bibinfo{note}{arXiv:1011.4553 (physics.ins-det)}.

\bibitem[{\citenamefont{Vergados}(2004)}]{JDV04}
\bibinfo{author}{\bibfnamefont{J.}~\bibnamefont{Vergados}},
  \bibinfo{journal}{J. Phys. G} \textbf{\bibinfo{volume}{30}},
  \bibinfo{pages}{1127} (\bibinfo{year}{2004}),
  \bibinfo{note}{[arXiv:hep-ph/0406134]}.

\bibitem[{\citenamefont{Vergados and Faessler}(2007)}]{VF07}
\bibinfo{author}{\bibfnamefont{J.}~\bibnamefont{Vergados}} \bibnamefont{and}
  \bibinfo{author}{\bibfnamefont{A.}~\bibnamefont{Faessler}},
  \bibinfo{journal}{Phys. Rev.} \textbf{\bibinfo{volume}{D 75}},
  \bibinfo{pages}{055007} (\bibinfo{year}{2007}).

\bibitem[{MeG()}]{MeGazSCH12}
\bibinfo{note}{J. Men?endez, D. Gazit, and A. Schwen, Spin-dependent WIMP
  scattering off nuclei, arXiv:1208.1094 [astroph.CO]}.

\bibitem[{Ver()}]{VergF12}
\bibinfo{note}{J. D. Vergados, Debris Flows in Direct Dark Matter Searches-The
  modulation effect, PRD (in press); arXiv:1202.3105 [hep-ph]}.

\bibitem[{\citenamefont{Vergados and Owen}(2007)}]{VEROW06}
\bibinfo{author}{\bibfnamefont{J.}~\bibnamefont{Vergados}} \bibnamefont{and}
  \bibinfo{author}{\bibfnamefont{D.}~\bibnamefont{Owen}},
  \bibinfo{journal}{Phys. Rev.} \textbf{\bibinfo{volume}{D 75}},
  \bibinfo{pages}{043503} (\bibinfo{year}{2007}).

\bibitem[{\citenamefont{Vergados}(2009)}]{JDV09}
\bibinfo{author}{\bibfnamefont{J.}~\bibnamefont{Vergados}},
  \bibinfo{journal}{Astronomical Journal} \textbf{\bibinfo{volume}{137}},
  \bibinfo{pages}{10} (\bibinfo{year}{2009}), \bibinfo{note}{[arXiv:0811.0382
  (astro-ph)]}.

\bibitem[{\citenamefont{Vergados}()}]{CHIOS07}
\bibinfo{author}{\bibfnamefont{J.~D.} \bibnamefont{Vergados}},
  \bibinfo{journal}{Lect. Notes Phys.}  (????).

\bibitem[{JEL()}]{JELLIS}
\bibinfo{note}{The Strange Spin of the Nucleon, J. Ellis and M. Karliner,
  hep-ph/9501280.}

\bibitem[{\citenamefont{Pittel and Vogel}(1992)}]{IJMPE}
\bibinfo{author}{\bibfnamefont{J.~E.~S.} \bibnamefont{Pittel}}
  \bibnamefont{and} \bibinfo{author}{\bibfnamefont{P.}~\bibnamefont{Vogel}},
  \bibinfo{journal}{J. Mod. Phys. E} \textbf{\bibinfo{volume}{1}},
  \bibinfo{pages}{1} (\bibinfo{year}{1992}).

\bibitem[{\citenamefont{Ressell and {\it et al.}}(1993)}]{Ressa}
\bibinfo{author}{\bibfnamefont{M.~T.} \bibnamefont{Ressell}} \bibnamefont{and}
  \bibinfo{author}{\bibnamefont{{\it et al.}}}, \bibinfo{journal}{Phys. Rev. D}
  \textbf{\bibinfo{volume}{48}}, \bibinfo{pages}{5519} (\bibinfo{year}{1993}).

\bibitem[{\citenamefont{Ressell and Dean}(1997)}]{Ressb}
\bibinfo{author}{\bibfnamefont{M.}~\bibnamefont{Ressell}} \bibnamefont{and}
  \bibinfo{author}{\bibfnamefont{D.~J.} \bibnamefont{Dean}},
  \bibinfo{journal}{Phys. Rev. C} \textbf{\bibinfo{volume}{56}},
  \bibinfo{pages}{535} (\bibinfo{year}{1997}).

\bibitem[{\citenamefont{J.Menendez et~al.}(2012)\citenamefont{J.Menendez,
  Gazit, and Schwenk}}]{MeGazSCH11}
\bibinfo{author}{\bibnamefont{J.Menendez}},
  \bibinfo{author}{\bibfnamefont{D.}~\bibnamefont{Gazit}}, \bibnamefont{and}
  \bibinfo{author}{\bibfnamefont{A.}~\bibnamefont{Schwenk}},
  \bibinfo{journal}{Phys. Rev. Lett.} \textbf{\bibinfo{volume}{107}},
  \bibinfo{pages}{62501} (\bibinfo{year}{2012}).

\bibitem[{\citenamefont{Cannoni}(2011)}]{Cannoni11}
\bibinfo{author}{\bibfnamefont{M.}~\bibnamefont{Cannoni}},
  \bibinfo{journal}{Phys. Rev. D} \textbf{\bibinfo{volume}{84}},
  \bibinfo{pages}{095017} (\bibinfo{year}{2011}),
  \bibinfo{note}{arXiv:1108.4337 (hep-ph)}.

\bibitem[{Can()}]{Cannoni12}
\bibinfo{note}{M. Cannoni, arXiv:1211.6050 (astro-phCO)}.

\bibitem[{\citenamefont{Cannoni et~al.}(2011)\citenamefont{Cannoni, Vergados,
  and Gomez}}]{CVG11}
\bibinfo{author}{\bibfnamefont{M.}~\bibnamefont{Cannoni}},
  \bibinfo{author}{\bibfnamefont{J.~D.} \bibnamefont{Vergados}},
  \bibnamefont{and} \bibinfo{author}{\bibfnamefont{M.~E.} \bibnamefont{Gomez}},
  \bibinfo{journal}{Phys. Rev. D} \textbf{\bibinfo{volume}{83}},
  \bibinfo{pages}{075010} (\bibinfo{year}{2011}),
  \bibinfo{note}{arXiv:1011.6108 (hep-ph)}.

\end{thebibliography}

\end{document}